\documentclass[amsmath,amssymb,twocolumn,showpacs,aps,prb,showkeys,superscriptaddress, 10pt]{revtex4-2} 

\usepackage{graphicx}
\usepackage{ifthen}
\usepackage{xcolor}
\usepackage{bm}
\usepackage{braket}
\usepackage{multirow}
\usepackage{hyperref}
\usepackage{soul}
\usepackage[normalem]{ulem}
\usepackage{multirow}
\hypersetup{
 pdfnewwindow=true, colorlinks=true,
 linkcolor=blue, anchorcolor=blue,
 citecolor=blue, filecolor=blue,
 menucolor=blue, urlcolor=blue}

\DeclareRobustCommand{\HITOSHI}[1]{\bgroup\color{olive}#1\egroup}

\DeclareMathSizes{10}{9.0}{6.5}{5.5} 

\def\bk{\mathbf{k}}
\def\br{{\bf r}}
\def\bkp{\mathbf{k^\prime}}

\def\bq{\mathbf{q}}
\def\bkq{\mathbf{k}+\mathbf{q}}
\def\bkp{\mathbf{k}+\mathbf{p}}
\def\bkqp{\mathbf{k}+\mathbf{q}+\mathbf{p}}
\def\bp{\mathbf{p}}
\def\o{\omega}
\def\a{\alpha}
\def\k{\kappa}
\def\bRp{{\bf R}_p}
\def\op{\omega^\prime}
\def\d{\delta}
\def\g{\gamma}
\def\ioj{i\o_j}
\def\iojp{i\o_{j^\prime}}
\def\iojdp{i\o_{j^{\prime \prime}}}
\def\iojtp{i\o_{j^{\prime \prime}}-i\o_{j}+i\o_{j^\prime}}
\def\etp{\ve''-\ve+\ve'}
\def\l{\lambda}
\def\D{\Delta}
\def\ve{\varepsilon}
\def\hS{\hat{\Sigma}}
\def\hG{\hat{G}}
\def\NF{N_{\mathrm{F}}}
\def\tg{\tilde{\gamma}}

\def\tc{T_{\rm c}}
\def\ef{\ve_{\rm F}}

\begin{document}

\title{Electron-phonon vertex correction effect in superconducting H$_3$S}

\author{Shashi B. Mishra}
\email{mshashi125@gmail.com}
\affiliation{Department of Physics, Applied Physics and Astronomy, Binghamton University-SUNY, Binghamton, New York 13902, USA}
\author{Hitoshi Mori}
\affiliation{Department of Physics, Applied Physics and Astronomy, Binghamton University-SUNY, Binghamton, New York 13902, USA}
\affiliation{Institute for Materials Research, Tohoku University, Sendai 980-8577, Japan}
\author{Elena R. Margine}
\email{rmargine@binghamton.edu}
\affiliation{Department of Physics, Applied Physics and Astronomy, Binghamton University-SUNY, Binghamton, New York 13902, USA}
\date{\today}

\begin{abstract}
The Migdal-Eliashberg (ME) formalism provides a reliable framework for describing phonon-mediated superconductivity in the adiabatic regime, where the electronic Fermi energy exceeds the characteristic phonon energy. In this work, we go beyond this limit by incorporating first-order vertex corrections to the electron-phonon (e-ph) interaction within the Eliashberg formalism and assess their impact on the superconducting properties of H$_3$S and Pb using first-principles calculations.
For H$_3$S, where the adiabatic assumption breaks down, we find that vertex corrections to the e-ph coupling are substantial. When combined with phonon anharmonicity and the energy dependence of the electronic density of states, the predicted critical temperature $\tc$ is in very good agreement with experimental observations. 
In contrast, for elemental Pb, where the adiabatic approximation remains valid, vertex corrections have a negligible effect, and the calculated $\tc$ and superconducting gap closely match the predictions of the standard ME formalism. 
These findings demonstrate the importance of non-adiabatic corrections in strongly coupled high-$\tc$ hydrides and establish a robust first-principles framework for accurately predicting superconducting properties across different regimes.
\end{abstract}

\maketitle

\section{\label{sec:intro}Introduction}

The Migdal-Eliashberg (ME) theory~\cite{Migdal1958,Eliashberg1960,Eliashberg1961} is a powerful many-body method for predicting superconducting properties and is widely regarded as the standard theory for conventional phonon-mediated superconductivity~\cite{Carbotte1990,Scalapino1966,Scalapino1969,Pickett1982,Allen1983,Marsiglio2008,Marsiglio2020,Pickett2023}. The theory, as formulated by Eliashberg~\cite{Eliashberg1960,Eliashberg1961}, extends the weak-coupling limit of the Bardeen-Cooper-Schrieffer (BCS) model~\cite{Bardeen1957} by incorporating the Green's function approach developed by Migdal for describing the electron-phonon (e-ph) interaction in the normal metallic state~\cite{Migdal1958}. Nowadays, first-principles simulations based on ME theory are performed routinely and at reasonable computational cost~\cite{Ponce2016,Sanna2020,Marini2024}, showing good agreement with experiments across a wide range of conventional superconductors, from simple metals to complex materials~\cite{Marques2005,Ryotaro2017,Sanna2018,Davydov2020,Flores2016,Pellegrini2024}. 
Nevertheless, ME theory relies on Migdal's theorem~\cite{Migdal1958}, which assumes that the dynamics of electrons and ions happen on two well-separated energy scales, in accordance with the adiabatic Born-Oppenheimer approximation~\cite{Born1927}. Migdal argued that vertex corrections to the bare e-ph scattering diagram can be neglected when the parameter $\l (\hbar \o_0/\ef) $ is small, where $\l$ is the effective e-ph coupling constant, $\o_0$ is the characteristic phonon frequency, and $\ef$ is the Fermi energy. Specifically, it was shown that each additional term in the perturbative expansion of the e-ph vertex is suppressed by at least a factor of the order $\hbar \o_0/\ef \sim \sqrt{m/M} \sim 10^{-2}$, where $m$ and $M$ are the electron and ion mass. The condition $\l (\hbar \o_0/\ef) \ll $ 1 is satisfied in most metallic systems since the adiabatic parameter $\hbar \o_0/\ef \ll $ 0, given that the electronic energy scale is of the order of several eV, whereas phonon energies are generally only a fraction of an eV. 
However, the validity of Migdal's theorem has been called into question for several classes of superconductors such as superhydrides, fullerides, and systems with low charge carrier densities, where the electronic Fermi energy is small~\cite{Jarlborg2016,Sano2016,Durajski2016,Gorkov2016,Gorkov2018,Roy2014,Gunnarsson1994,Cappelluti2000,Gastiasoro2019,Hu2022,Gorkov2016b,Zhang2024}. A natural generalization of the Eliashberg theory to the non-adiabatic regime can be achieved through a perturbative diagrammatic approach in the expansion parameter $\l (\hbar \o_0/\ef)$ ~\cite{Pietronero1992,Kostur1993,Kostur1994,Grimaldi1995,Pietronero1995,Grimaldi1995b,Botti2002,Pietronero2006,Velasco2025,Esterlis2018}. 
This perturbative scheme is expected to remain applicable as long as $\l$ is not larger than a critical value (of order 1) beyond which the system undergoes lattice transitions, leading to the formation of charge-density-waves or polaronic and/or bipolaronic states \cite{Alexandrov2001,Esterlis2018,Chubukov2020,Cappelluti2023}. 
In recent years, there has been growing interest in understanding the impact of higher-order e-ph processes on key materials properties (e.g., spectral functions, band-gap renormalization, polaron formation, carrier mobility, or superconducting pairing) but studies that move beyond leading-order approximations remain rare~\cite{Sano2016,Schrodi2020,Lee2020,Hu2022,Bianco2023,Houtput2025,Zappacosta2025,Lihm2025a,Lihm2025b}. This scarcity is primarily due to the rapid increase in computational complexity, as even incorporating second-order scattering events is highly nontrivial. Nonetheless, promising progress has been made, with several recent studies achieving such treatments fully from first principles~\cite{Lee2020,Bianco2023,Houtput2025}. On the superconductivity front, 
current approaches to non-adiabatic effects within the Eliashberg framework have  relied on simplifying approximations. For example, in the anisotropic case, model Hamiltonians for the electronic structure, combined with an isotropic Einstein phonon spectrum and a single e-ph parameter, have been employed to solve the non-adiabatic Eliashberg equations~\cite{Schrodi2020}. In the isotropic case,  besides replacing the momentum dependence in all kernels with Fermi-surface averages, further simplifications have been introduced, such as factorizing the lowest-order vertex correction kernel into a product of two e-ph interaction kernels~\cite{Freericks1997,Durajski2016,Schrodi2020}.
In this work, we develop and implement a first-principles formalism to compute second-order e-ph vertex corrections within the isotropic Eliashberg framework~\cite{Eliashberg1960,Eliashberg1961}, extending beyond the conventional ME approximation~\cite{Migdal1958,Allen1983}. 
The main challenge in implementing this general treatment lies in the significantly increased computational cost, driven by two key factors. First, evaluating the vertex-corrected Eliashberg spectral function requires the computation of four distinct e-ph matrix elements associated with two phonons. Second, incorporating the resulting vertex-corrected e-ph interaction kernel into the Eliashberg equations introduces additional summations over Matsubara frequencies and integrations over electronic energies. Our methodology is implemented in the {\textsc EPW} code and supports both the full-bandwidth ({\small FBW}) and Fermi-surface restricted ({\small FSR}) approaches~\cite{Margine2013,Lee2023,Lucrezi2024}. The {\small FBW} scheme retains the full energy dependence of the density of states (DOS), enabling the inclusion of e-ph scattering processes away from the Fermi level ($\ef$), while {\small FSR} assumes a constant DOS near $\ef$. 
To demonstrate the capabilities of our non-adiabatic methodology, we investigate the effect of vertex corrections on the superconducting gap and critical temperature ($\tc$) of H$_3$S. This high-$\tc$ superconductor serves as a prime example of a system where the adiabatic limit breaks down~\cite{Jarlborg2016,Gorkov2018}. At 200~GPa, a flat-band feature near $\ef$ gives rise to a van Hove singularities (vHs), leading to a small effective Fermi energy. Moreover, the characteristic phonon energy in H$_3$S exceeds 200~meV, yielding an adiabatic parameter of $\hbar \o_0/\ef \approx$ 5~\cite{Gorkov2018}. We find that the e-ph coupling constant with vertex corrections ($\lambda^{\rm V}$) remains a significant fraction of the adiabatic coupling strength ($\lambda$), highlighting the persistence of strong coupling beyond the adiabatic regime in systems where the adiabatic limit breaks down, as is the case for many recently studied high-$\tc$ hydrides~\cite{Gorkov2018,Pickett2023,Hilleke2022}.
Apart from vertex corrections, another critical factor in H$_3$S is the breakdown of the harmonic approximation, driven by hydrogen's low mass and significant quantum fluctuations~\cite{Errea2015,Sano2016,Taureau2024}. These anharmonic effects renormalize phonon frequencies, particularly hydrogen-related modes, leading to strong phonon hardening and a consequent reduction in both the e-ph coupling and critical temperature. To examine the interplay between anharmonic and non-adiabatic effects, we compute the e-ph interactions and superconducting properties of H$_3$S by solving the Eliashberg equations in both the adiabatic and non-adiabatic regimes, with and without anharmonic phonons. Our results show that when both anharmonic and non-adiabatic effects are considered together, the resulting $\tc$ falls within the experimental range. 
Finally, we analyze the impact of vertex corrections in Pb, where the adiabatic limit is generally considered valid. In this case, including vertex corrections results in a $\tc$ nearly identical to that predicted within the adiabatic framework, confirming their negligible effect in systems where the Migdal approximation holds.

This paper is organized as follows. In Sec.~\ref{sec:derivation}, we present the non-adiabatic Eliashberg theory and describe our \textit{ab initio} implementation with lowest-order vertex corrections. Computational details and results for H$_3$S and Pb, both with and without vertex corrections, are provided in Secs.~\ref{sec:methods} and \ref{sec:results}, respectively. Finally, Sec.~\ref{sec:summary} summarizes our findings. Additional details, including diagrammatic representation of e-ph self-energy, derivation of the isotropic non-adiabatic self-energy expression, symmetry reduction in the vertex-corrected spectral function, derivation of the isotropic non-adiabatic Eliashberg equations for both the {\small FBW} and {\small FSR} approaches, summation procedure over the Matsubara frequencies as well as convergence tests and additional results for H$_3$S with harmonic and anharmonic phonons, are included in the Supplementary Information~\cite{SI}. 

\section{Theory}\label{sec:derivation}

In this section, we review the theoretical background underlying phonon-mediated superconductivity, focusing on ME formalism and its extension to include the lowest-order vertex corrections. To solve the Eliashberg equations, we employ the {\small FBW} approach, which accounts for the full energy dependence of the electronic DOS. This is essential for accurately capturing the effects of sharp structures in the DOS, such as vHs, and allows for calculations with both fixed and self-consistently updated chemical potential ($\mu_{\rm F}$). We also present simplified expressions based on the constant DOS {\small FSR} approach. 

\subsection{Electron self-energy with electron-phonon vertex corrections}

Within the Nambu formalism~\cite{Nambu1960}, the Hamiltonian of an electron-phonon interacting system can be written as  
\begin{align}\label{Ham1}
\hat{H} & =  \hat{H}_e + \hat{H}_{ph} + \hat{H}_{e\text{-}e} + \hat{H}_{e\text{-}ph},
\end{align}
where
\begin{align} 
\hat{H}_e &= \sum_{n \bk} \ve_{n \bk} \hat{\Psi}_{n \bk}^\dagger \hat{\tau}_3 \hat{\Psi}_{n \bk} \label{eq:ele}\\
\hat{H}_{ph} &= \sum_{\bq \nu } \hbar \o_{\bq \nu} \! \left(\hat{a}_{\bq \nu}^\dagger \hat{a}_{\bq \nu} + \frac{1}{2}\right) 
\label{eq:ph}
\end{align}
\begin{align}
\hat{H}_{e-e} &= \frac{1}{2} \sum_{n m} \sum_{\bk \bk'}  W_{n\bk,m\bk'} \! 
\left(\hat{\Psi}_{m\bk'}^\dagger \hat{\tau}_3 \hat{\Psi}_{n\bk}\right) \! \! 
\left(\hat{\Psi}_{m-\bk'}^\dagger \hat{\tau}_3 \hat{\Psi}_{n-\bk}\right)  
\label{eq:e-e}
\end{align}
\begin{align}
\hat{H}_{e\text{-}ph} &= \frac{1}{\sqrt{N_p}}\sum_{nm \nu} \sum_{\bk \bq}  g_{m\bkq,n\bk}^{\bq\nu} \! 
\left(\hat{a}_{\bq \nu} + \hat{a}_{-\bq \nu}^\dagger \right) \! 
\Psi_{m\bkq}^\dagger \hat{\tau}_3 \Psi_{n\bk}. \label{eq:e-ph}
\end{align}
Here, $\hat{\tau}_i$ $(i=0,1,2,3)$ are the Pauli matrices. Each of the terms is described below.
The first term represents the electron Hamiltonian, where $\hat{\Psi}_{n\bk}$ is a two-component electron field operator
\begin{equation}
    \hat{\Psi}_{n\bk} = \begin{pmatrix}
        \hat{c}_{n\bk\uparrow} \\
        \hat{c}^\dagger_{n-\bk\downarrow}
    \end{pmatrix},
    \, \hat{\Psi}^\dagger_{n\bk} = \begin{pmatrix}
        \hat{c}^\dagger_{n\bk\uparrow} \,\, \hat{c}_{n-\bk\downarrow}
    \end{pmatrix},
\end{equation}
with $\hat{c}^\dagger_{n\bk\sigma}$ and $\hat{c}_{n\bk\sigma}$ are the creation and annihilation operators for an electronic state with momentum $\bk$, band index $n$, spin $\sigma$, and energy $\ve_{n \bk}$. The second term describes the phonon Hamiltonian, where $\o_{\bq \nu}$ is the frequency of a phonon mode with momentum $\bq$ and branch index $\nu$, and $a^\dagger_{\bq \nu}$ and $a_{\bq \nu}$ are the bosonic creation and annihilation operators. 
The anharmonic terms, when relevant, are incorporated into Eq.~\eqref{eq:ph} through additional contributions derived from the self-consistent phonon theory, which replaces the true double-well potential with an effective temperature-dependent harmonic potential that minimizes the free energy, as discussed in detail in Ref.~\cite{Zacharias2023}. 
The third term accounts for the Coulomb contribution, where $W_{n\bk,m\bkq}$ are the scattering matrix elements between electrons. 

\begin{figure*}[!ht]
\centering
\includegraphics[width=0.85\textwidth]{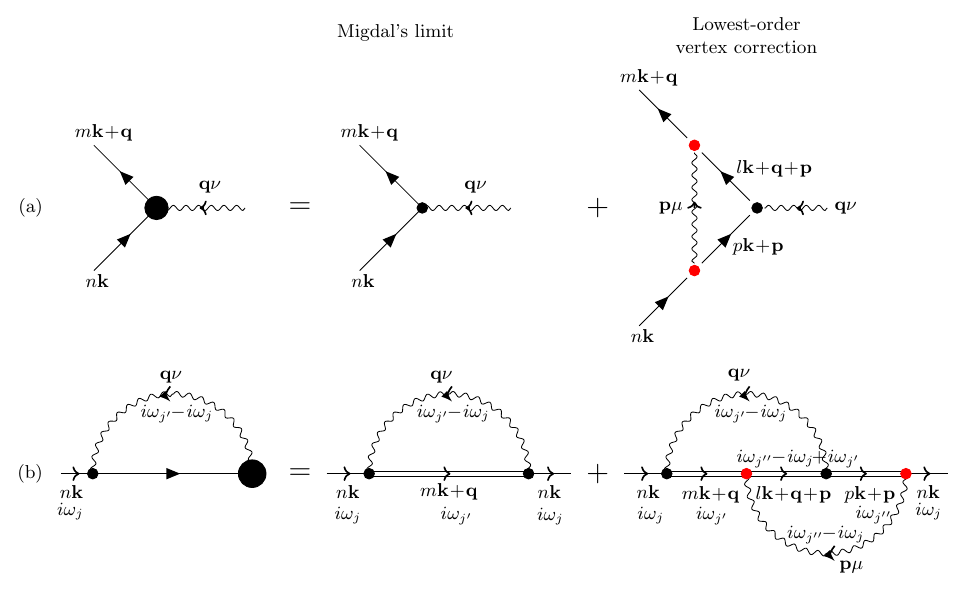}
\caption{\label{fig-1}First and second-order Feynman diagrams illustrating the (a) e-ph vertex and (b) the electron self-energy contributions arising from e-ph interactions. In these diagrams, straight lines represent bare electron Green’s functions,
straight double lines represent fully dressed electron Green’s functions, and wavy lines denote bare phonon propagators. The second phonon-mediated e-ph vertex is indicated by red dots.}
\end{figure*}

The fourth term describes the e-ph interaction, as illustrated in Fig.~\ref{fig-1}(a), where $N_p$ denotes the number of unit cells in the Born-von K\'arm\'an (BvK) supercell. This term corresponds to the conventional linear coupling between an electron and a phonon, commonly referred to as the Fan-Migdal term. Higher-order contributions to $\hat{H}_{e-ph}$, known as nonlinear e-ph interactions, could also influence materials properties~\cite{Heid1992,Dee2020,Bianco2023,Houtput2025,Zappacosta2025,Lihm2025a,Lihm2025b}, but are not considered in this work.
The linear e-ph interaction term is governed by the matrix element $g_{m\bkq,n\bk}^{\bq\nu}$, which quantifies the scattering between electronic states $n\bk$ and $m\bk\!+\!\bq$ mediated by a phonon with wavevector $\bq$ and branch index $\nu$. This matrix element is defined as~\cite{Giustino2007,Giustino2017} 
\begin{align}{\label{eq:g-exp}}
    g_{m\bkq,n \bk}^{\bq \nu} &= 
     \bra{u_{m\bk+\bq}} \Delta_{\bq\nu}v^{\rm KS} \ket{u_{n\bk}}_{\rm uc},
\end{align}
where $u_{n\bk}$ is the Bloch-periodic part of the Kohn-Sham (KS) electron wavefunction, and the integral is performed over the unit cell (uc). The derivative of the self-consistent potential is given by
\begin{align}\label{eq:dv}
\Delta_{\bq\nu}v^{\rm KS} &\equiv  e^{-i\bq \cdot \br} \Delta_{\bq\nu}V^{\rm KS} \nonumber \\
&= \! \! \sum_{\kappa\a \bRp} \! \! e^{-i\bq \cdot (\br - \bRp)} 
\sqrt{\frac{\hbar}{2M_\kappa\o_{\bq\nu}}} e_{\k\a,\nu}(\bq) 
\frac{\partial V^{\rm KS}(\br)}{\partial \tau_{\k\a p}}. 
\end{align}
Here, $\kappa$ labels atoms in the unit cell with $M_\kappa$ as the atomic mass, and $\a$ denotes Cartesian directions. The index $p$ labels unit cells with BvK boundary conditions, $\tau_{\k \a p}$ is the position of atom $\kappa$ in the unit cell $p$, $\bf{R}_p$ is the lattice vector identifying the unit cell $p$, and $e_{\kappa \a, \nu}(\bq)$ is the eigenvector corresponding to atom $\kappa$ in the Cartesian direction $\a$ for a collective phonon mode $\bq \nu$. 
The differential $\partial V^{\rm KS}(\br) / \partial \tau_{\k\a p}$ is computed using density functional perturbation theory (DFPT)~\cite{Baroni2001}. In the presence of anharmonicity, the phonon modes are no longer exact eigenmodes of a harmonic potential, making $\Delta_{\bq\nu}v^{\rm KS}$ formally ill-defined. To incorporate anharmonic effects within Eq.~\eqref{eq:dv}, we retain the harmonic DFPT-computed potential derivatives while replacing the phonon eigenvectors and frequencies with their anharmonically renormalized counterparts, obtained by diagonalizing the dynamical matrix obtained from the special displacement method~\cite{Zacharias2023}. 
Using Nambu's notation, we define the finite-temperature electron Green's function as a \(2 \times 2\) matrix~\cite{Nambu1960}:
\begin{align}\label{G}
\hG_{n \bk}(\tau) &= - \langle \hat{T}_{\tau} \hat{\Psi}_{n\bk} (\tau) \hat{\Psi}^\dagger_{n\bk} (0) \rangle \nonumber \\
&= - \! \begin{pmatrix}
\langle \hat{T}_{\tau} \hat{c}_{n\bk\uparrow}(\tau) \hat{c}^\dagger_{n\bk\uparrow}(0) \rangle & 
\langle \hat{T}_{\tau} \hat{c}_{n\bk\uparrow}(\tau) \hat{c}_{n-\bk\downarrow}(0) \rangle \\
\langle \hat{T}_{\tau} \hat{c}^\dagger_{n-\bk\downarrow}(\tau) \hat{c}^\dagger_{n\bk\uparrow}(0) \rangle & 
\langle \hat{T}_{\tau} \hat{c}^\dagger_{n-\bk\downarrow}(\tau) \hat{c}_{n-\bk\downarrow}(0) \rangle
\end{pmatrix},
\end{align}
where \(\hat{T}_{\tau}\) denotes Wick’s time-ordering operator in imaginary time, and \(\langle \cdots \rangle\) indicates the grand-canonical thermodynamic average. The diagonal elements correspond to the normal Green's functions for spin-up and spin-down electrons, describing single-particle electronic excitations. The off-diagonal elements are the anomalous Green's functions, introduced by Gor'kov~\cite{Gorkov1958}, which describe superconducting Cooper pairs. An important feature of the Nambu-Gor'kov formalism is that, in this matrix representation, the familiar Feynman-Dyson rules of many-body perturbation theory remain valid~\cite{Schrieffer1999}.

The study of the superconducting state involves the determination
of the matrix Green’s function in Eq.~\eqref{G}. In Matsubara space, this matrix Green's function obeys the Dyson equation~\cite{Migdal1958,Mahan1993}:
\begin{equation}\label{G-1}
\left[\hG_{n \bk}(\ioj)\right]^{-1}  = \left[\hG_{n \bk}^{0}(\ioj) \right]^{-1} - \,\hS_{n \bk}(\ioj), 
\end{equation}
where $\hG_{n \bk}^{0}(\ioj)$ is the non-interacting Green's function
\begin{equation}\label{G-0}
\left[\hG_{n \bk}^{0}(\ioj)\right]^{-1} = i\o_{j}\hat{\tau}_0 - (\ve_{n \bk} - \mu_{\rm F})\hat{\tau}_3,
\end{equation}
and $\hS_{n \bk}(\ioj)$  is the electron self-energy including contributions arising from the e-ph and Coulomb interactions
\begin{equation}
\hS_{n \bk}(\ioj) = \hS^{\rm ep}_{n \bk}(\ioj) + \hS^{\rm c}_{n \bk}(\ioj).   
\end{equation}
In these expressions, $\ioj= i (2j+1) \pi k_{\rm B}T$ denotes the fermionic Matsubara frequency, where $j$ is an integer, $T$ is the temperature, and $k_{\rm B}$ is the Boltzmann constant. 

Following the standard practice,  the Coulomb part of the electron self-energy is defined within the \textit{GW} approximation~\cite{Hedin1965,Hybertsen1986} as 
\begin{equation}
\label{Sigma_c}
	\hS^{\rm c}_{n \bk}(i\o_j)
	= -k_{\rm B} T \sum_{m \bk'}\sum_{j'} W_{n \bk, m \bk'}(\ioj-\iojp) \hat{\tau}_3 \hG_{m \bk'}(\iojp)
	\hat{\tau}_3  ,
\end{equation}
where $W_{n \bk, m \bk'}(\ioj-\iojp)$ is the dynamically screened Coulomb interaction. Assuming a static screened Coulomb interaction, Eq.~\eqref{Sigma_c} simplifies to
\begin{equation}
\label{Sigma_c2}
	\hS^{\rm c}_{n \bk}(i\o_j)
	=  -k_{\rm B} T \sum_{m \bk'}\sum_{j'} W_{n \bk, m \bk'} \hat{\tau}_3 \hG^{\rm od}_{m \bk'}(\iojp)
	\hat{\tau}_3  ,
\end{equation}
where only the off-diagonal components of the Green’s function $\hG^{\rm od}_{n \bk}(i\o_j)$ are retained to avoid double counting Coulomb effects, since the one-particle eigenenergy $\ve_{n\bk}$ in $\hG_{n \bk}^{0}(\ioj)$ already includes part of the Coulomb interaction~\cite{Allen1983}. 

Recent methodological advancements have enabled a fully \textit{ab initio} treatment of Coulomb interactions within Eliashberg theory~\cite{Ryotaro2017,Sanna2018,Davydov2020,Pellegrini2023,Mori2024,kogler2025}. However, explicitly incorporating the full Coulomb kernel $W_{n \bk, m \bk'}$ in Eq.~\eqref{Sigma_c2} remains computationally challenging. As a result, this interaction is typically approximated using the semi-empirical Morel-Anderson $\mu_{\rm c}^*$ parameter~\cite{Morel1962,Scalapino1966}, defined as
\begin{align}\label{eq:muc}
    \mu_c^* = \frac{\mu_c}{1+\mu_c \, \rm{ln} \left(\frac{\ve_{el}}{\hbar \o_0}\right)},
\end{align}
where $\mu_c$ is the Fermi-surface averaged Coulomb interaction, $\o_0$ is a characteristic phonon frequency, and $\ve_{el}$ is the electronic bandwidth. Within the Eliashberg formalism, the corresponding phonon energy scale is given by the Matsubara frequency cutoff $\o_c$, thus $\mu_c^*$ must be replaced by $\mu^*_{\rm E}$, which are related via~\cite{Allen1975,Pellegrini2024} 
\begin{align}\label{eq:muE}
    \frac{1}{\mu^*_{\rm E}} = \frac{1}{\mu_c^*} + \rm{ln} \left(\frac{\o_0}{\o_c}\right).
\end{align}
This approximation is used in the current study. 
For the phonon part, the central approximation consists of retaining only the first-order e-ph scattering diagram, shown in Fig.~\ref{fig-1}(b), for $\hS^{\rm ep}_{n \bk}(\ioj)$, in line with Migdal's theorem~\cite{Migdal1958}. However, as discussed in Sec.~\ref{sec:intro}, this adiabatic approximation fails in several classes of superconductors, necessitating the inclusion of higher-order vertex corrections to the e-ph interaction~\cite{Allen1983,Pietronero1995,Botti2002,Pietronero2006,Cappelluti2006}, which forms the main focus of this work. 

A detailed diagrammatic analysis, presented in Fig.~S1 of the supplementary Information~\cite{SI}, demonstrates that next-order self-energy diagrams, such as the rainbow and parallel bubble diagrams, correspond to Fermi-surface-confined processes whose contributions are already implicitly accounted for in the Migdal approximation~\cite{Allen1983,Heid2017,Schrodi2020}. In contrast, the lowest-order vertex correction diagram in Fig.~\ref{fig-1}(b) captures non-adiabatic effects by involving intermediate electronic states that lie off-shell from the Fermi-surface, as illustrated in the corresponding scattering diagram shown in Fig.~S1(h)~\cite{SI}. In this work, we only include the lowest-order vertex correction. The explicit treatment of higher-order multi-phonon processes is deferred to future work, motivated by recent analytical toy model results indicating that the rainbow diagram can also influence the superconducting gap~\cite{Zappacosta2025}.
The electron self-energy $\hS^{\rm ep}_{n \bf{k}}(i\o_j)$, incorporating both adiabatic $\mathrm{(A)}$ and non-adiabatic $\mathrm{(NA)}$ effects, is given by: 
\begin{align}\label{total}
\hS^{\rm ep}_{n \bk}(i\o_j) = \Sigma_{n \bf{k}}^{\mathrm{(A)}}(i\o_j)
+ \Sigma_{n \bf{k}}^{\mathrm{(NA)}} (i\o_j).
\end{align}
The adiabatic self-energy $\hS_{n \bk}^{\mathrm{(A)}}(\ioj)$, corresponding to the first-order Feynman diagram in Fig.~\ref{fig-1}, is expressed as
\begin{align}\label{selfA}
\hS_{n \bk}^{\mathrm{(A)}}(\ioj) &=
 -k_{\rm B}T \sum_{m \bq}  \sum_{\nu j'} 
|g_{m\bkq,n \bk}^{\bq \nu}|^2  D_{\bq \nu }(\ioj-\iojp) \nonumber \\
&\quad \times \hat{\tau}_3 \hG_{m \bkq}(\iojp) \hat{\tau}_3,
\end{align}
where $g_{m\bkq,n \bk}^{\bq \nu}$ is the screened e-ph coupling matrix element.  
Since $\hat{H}_{e-ph}$ is Hermitian, the e-ph matrix element satisfy $g_{m\bkq,n\bk}^{\bq \nu} = \left[g_{n\bk,m\bkq}^{-\bq\nu}\right]^*$.

Following the standard practice~\cite{Allen1983}, we replace the full phonon propagator $D_{\bq \nu}(\ioj-\iojp)$ by the bare phonon Green's function
\begin{align}\label{D0}
D_{\bq \nu}^0(\ioj-\iojp) &= -\frac{2\o_{\bq \nu}}{(\o_j - \o_{j'})^2 + \o_{\bq \nu}^2}.
\end{align}

We define the anisotropic e-ph interaction kernel as
\begin{align}\label{l-A-aniso}
\l_{n \bk, m \bkq}(\ioj-\iojp) 
& = \int_{0}^{\infty} d\o \frac{2\o}{(\o_j - \o_{j'})^2 + \o^2} \nonumber \\
& \quad \quad \times \alpha^2F_{n \bk, m\bkq}(\o),
\end{align}
where the anisotropic Eliashberg spectral function is given by
\begin{align}\label{a2f-aniso}
\alpha^2F_{n\bk, m\bkq}(\o) 
&= \NF \sum_{\nu} |g_{m\bkq,n \bk}^{\bq \nu}|^2 \d(\o-\o_{\bq \nu }),
\end{align}
with $\NF$ as the DOS per spin at $\ef$. Substituting Eqs.~\eqref{D0} $-$\eqref{a2f-aniso} into the self-energy expression, Eq.~\eqref{selfA} takes the compact form
\begin{align}\label{selfA-1}
\hS_{n \bk}^{\mathrm{(A)}}(\ioj) \! &=
k_{\rm B} T \sum_{m\bq j'} \! \!
\frac{\l_{n \bk, m \bkq}(\ioj-\iojp)}{\NF}
\hat{\tau}_3 \hG_{m \bkq}(\iojp) \hat{\tau}_3.
\end{align}
The non-adiabatic self-energy $\hS_{n \bk}^{\mathrm{(NA)}}(\ioj)$, corresponding to the second-order Feynman diagram in Fig.~\ref{fig-1}(b), is given by
\begin{widetext}
\begin{align}\label{self-NA}
\hS_{n \bk}^{\mathrm{(NA)}}(\ioj) &= (k_{\rm B}T)^2 \sum_{mlp} 
\sum_{\bq \bp} 
\sum_{\nu \mu}  
\sum_{j'j''} 
g_{m\bkq,n\bk}^{\bq \nu}
g_{l\bkqp, m\bkq}^{\bp \mu}
g_{p\bkp, l\bkqp}^{-\bq \mu}
g_{n\bk, p\bkp}^{-\bp \mu} 
D_{\bq \nu} (\ioj-\iojp) 
D_{\bp \mu} (\ioj-\iojdp) \nonumber \\
&\quad  \quad \quad \quad \times  
\hat{\tau}_3\hG_{m \bkq}(\iojp)\hat{\tau}_3  
\hG_{l\bkqp}(\iojtp) 
\hat{\tau}_3\hG_{p \bkp}(\iojdp)\hat{\tau}_3.
\end{align}
Here, each matrix element represents an e-ph scattering vertex, associated with either phonon absorption ($\bq \nu$ or  $\bp \mu$) or phonon emission ($-\bq \nu$ or $-\bp \mu$). The expression captures a two-phonon scattering process involving four distinct interactions, each requiring explicit evaluation of the corresponding matrix element.

Analogous to the adiabatic case, we substitute the phonon propagators with their bare counterparts and introduce the anisotropic first-order vertex correction to the e-ph coupling via a double spectral representation
\begin{align} \label{l-NA-aniso}
\l^{\rm V}_{ n\bk, m\bkq, l\bkqp, p\bkp}(\ioj -\iojp, \ioj -\iojdp)
&= \int_{0}^{\infty} d\o \int_{0}^{\infty} d\op
\frac{2\o} {(\o_j - \o_{j'})^2 + \o^2} \times \frac{2\op} {(\o_j - \o_{j''})^2 + \op{^2}} \nonumber \\
&\quad \times \a^2F^{\rm V}_{n\bk, m\bkq, l\bkqp, p\bkp}(\o, \op),
\end{align}
where the anisotropic first-order vertex correction to the Eliashberg spectral function is expressed as follows
\begin{align} \label{a2f-NA}
\a^2F^{\rm V}_{n\bk, m\bkq, l\bkqp, p\bkp}(\o, \op) &=  \NF^2 \sum_{\nu\mu}
g_{m\bkq,n\bk}^{\bq \nu}
g_{l\bkqp, m\bkq}^{\bp \mu}
g_{p\bkp, l\bkqp}^{-\bq \mu}
g_{n\bk, p\bkp}^{-\bp \mu}  
\d(\o-\o_{\bq \nu}) \d(\op-\o_{\bp \mu}).
\end{align}
Using Eqs.~\eqref{l-NA-aniso} and \eqref{a2f-NA}, the non-adiabatic electron self-energy can be expressed in a compact form as
\begin{align}\label{self-NA-1}
\hS_{n \bk}^{\mathrm{(NA)}}(\ioj)&=
\frac{(k_{\rm B}T)^2}{\NF^2} 
\sum_{mlp} \sum_{\bq \bp} \sum_{j'j''}
\hat{\tau}_3\hG_{m \bkq}(\iojp)\hat{\tau}_3  
\hG_{l\bkqp}(\iojtp) 
\hat{\tau}_3\hG_{p \bkp}(\iojdp)\hat{\tau}_3
\nonumber \\
& \quad \quad  \quad  \quad \quad  \quad  \quad \times 
\l^{\rm V}_{n\bk, m\bkq, l\bkqp, p\bkp}(\ioj -\iojp, \ioj -\iojdp).
\end{align}
\end{widetext}

\subsection{Isotropic approximation to electron self-energy}

Accounting for the anisotropy in the non-adiabatic self-energy is computationally demanding, as Eq.~\eqref{self-NA} involves Brillouin zone integrals over two crystal momenta. To date, the most advanced calculations of superconducting properties with anisotropic vertex corrections have relied on several approximations, which include using model Hamiltonians for electronic states, an isotropic Einstein phonon spectrum, and a single e-ph coupling parameter~\cite{Schrodi2020}. However, for most materials, except for a few notable layered systems such as MgB$_2$-type superconductors~\cite{Szabo2001,Margine2013,Tomassetti2024_2}, graphite intercalation compounds~\cite{Margine2016,Mishra2024}, and transition metal dichalcogenides~\cite{Heil2017,Das2023}, the effect of this full momentum dependence is weak and an isotropic treatment can be employed~\cite{Davydov2020,Pellegrini2024}. This approximation consists of replacing the $\bk$-dependent quantities with their Fermi-surface averages. A detailed derivation is provided in Sec.~II of the Supplementary Information~\cite{SI}.
 
Within the isotropic approximation, the adiabatic self-energy takes the following form
\begin{align}\label{self-A-iso} 
    \hS^{\mathrm{(A)}}(\ioj) 
    &= \frac{k_{\rm B}T}{\NF} \sum_{j'} \l(\ioj-\iojp)
    \! \int_{-\infty}^{\infty}\! \!  \! d\ve' 
    N(\ve')  \hat{\tau}_3 \hG(\ve',\iojp) \hat{\tau}_3, 
\end{align}
where $\l(\ioj-\iojp)$ is the isotropic e-ph kernel, expressed as
\begin{align}\label{l-A-iso}
    \l(\ioj-\iojp) 
    = \int_{0}^{\infty} 
    d\o \frac{2\o}{(\o_j - \o_{j'})^2 + \o^2}
    \, \alpha^2F(\o).
\end{align}
The kernel is an even function of the Matsubara frequency and attains its maximum at $\o_j - \o_{j'}=0$
\begin{align}
    \l = 2 \int_{0}^{\infty} 
    d\o \frac{\alpha^2F(\o)}{\o},
\end{align}
where $\l$ denotes the isotropic e-ph coupling strength.
The isotropic Eliashberg spectral function $\alpha^2F(\o)$ represents the double Fermi-surface average of $\alpha^2F_{n\bk, m\bkq}(\o)$ and is given by
\begin{align}\label{a2f-A-iso}
\alpha^2F(\o) 
  & =  \frac{1}{\NF} 
  \sum_{nm} \sum_{\bk \bq}
  \d(\ve_{n \bk}-\ef)
  \d(\ve_{m \bkq}-\ef) 
  \nonumber \\
  &\quad \quad \times \sum_{\nu} |g_{m\bkq,n \bk}^{\bq \nu}|^2 
  \d(\o-\o_{\bq \nu}),
\end{align}
Following similar steps, the isotropic non-adiabatic self-energy associated with the lowest-order vertex correction can be simplified as
\begin{widetext}
\begin{align} \label{self-NA-iso-1}
\hS^{\mathrm{(NA)}}(\ioj) =  
\frac{(k_{\rm B}T)^2}{\NF^2} & \sum_{j'j''} 
\l^{\rm V}(\ioj -\iojp, \ioj -\iojdp)
\int_{-\infty}^{\infty}\! d\ve' N(\ve') 
\int_{-\infty}^{\infty}\! d(\etp) N(\etp)
\int_{-\infty}^{\infty}\! d\ve'' N(\ve'')
\nonumber \\
& \hspace{0.1cm}\times 
\hat{\tau}_3\hG(\ve',\iojp)\hat{\tau}_3\hG(\etp,\iojtp) \hat{\tau}_3\hG(\ve'',\iojdp)\hat{\tau}_3,
\end{align}

\noindent where the isotropic vertex-corrected e-ph kernel
\begin{align}\label{NA-lambda-iso}
\l^{\rm V}(\ioj -\iojp, \ioj -\iojdp)
& = \int_{0}^{\infty}
d\o \frac{2\o}{(\o_j - \o_{j'})^2 + \o^2}
\int_{0}^{\infty} 
d\op \frac{2\op}{(\o_j - \o_{j''})^2 + \op{^2}}
\a^2F^{\rm V}(\o, \op).
\end{align}
The vertex-corrected kernel attains its maximum at $\ioj -\iojp=0 \text{ and } \ioj -\iojdp=0$, and reduces to
\begin{align}\label{lambda-ver}
\l^{\rm V}
& = 4 \int_{0}^{\infty} d\o \int_{0}^{\infty} d\op 
\frac{\a^2F^{\rm V}(\o, \op)}{\o \,  \op}.
\end{align}
Here, $\l^{\rm V}$ is called the vertex-corrected e-ph coupling strength and is a dimensionless measure of the average e-ph coupling strength arising from the non-adiabatic vertex corrections.
The isotropic vertex-corrected Eliashberg spectral function $\a^2F^{\rm V}(\o, \op)$ is obtained by performing the Fermi-surface average as follows 
\begin{align}\label{a2f-NA-iso}
\a^2F^{\rm V}(\o, \op)&= 
\frac{1}{C(0)}
    \sum_{n mlp} \sum_{\bk \bq \bp} 
  \d(\ve_{n \bk}-\ef) 
  \d(\ve_{m \bkq}-\ef)
 \d(\ve_{l \bkqp}-\ef) 
 \d(\ve_{p \bkp}-\ef)  
 \a^2F^{\rm V}_{n \bk, m\bkp, l\bkqp, p\bkp}(\o, \op).
\end{align}
\end{widetext}
\textnormal{Here, the normalization factor $ C(0)$ corresponds to the total weight of all four-fold combinations of states at $\ef$}
\begin{align}\label{C-NA-iso}
C(0) &= 
\sum_{n mlp} \sum_{\bk \bq \bp}   
\d(\ve_{n \bk}-\ef) 
  \d(\ve_{m \bkq}-\ef)
  \nonumber \\
  &\quad \quad \times
 \d(\ve_{l \bkqp}-\ef) 
 \d(\ve_{p \bkp}-\ef).
\end{align}

Computing Eq.~\eqref{a2f-NA-iso} has remained a longstanding challenge, even in the isotropic framework, due to the requirement of evaluating four e-ph matrix elements involving two different phonons. To date, additional approximations have been employed, such as estimating the density factor $C(0)$ based on the free-electron model~\cite{Freericks1997,Durajski2016}, and expressing the vertex-corrected e-ph term $\l^{\rm V}(\ioj -\iojp, \ioj -\iojdp)$ as a product of two e-ph vertices $\l(\ioj-\iojp)\l(\ioj-\iojdp)$~\cite{Freericks1997,Schrodi2020}. In this work, we compute this term without relying on these simplifications. We exploit the symmetry under phonon exchange $(\bq, \nu, \o) \leftrightarrow (\bp, \mu, \op)$ to reduce the computational cost. This symmetry, evident from the Feynman diagrams in Fig.~\ref{fig-1}(b), enables us to express $\a^2F^{\rm V}(\o, \op)$ in terms of the real part of the four e–ph matrix elements product. A detailed derivation is provided in Sec.~III of the Supplementary Information~\cite{SI}.

\subsection{Isotropic non-adiabatic full-bandwidth Eliashberg equations}

To obtain the Eliashberg equations, the standard procedure is to express the total self-energy on the basis of Pauli matrices in terms of three scalar functions~\cite{Carbotte1990,Scalapino1966,Allen1983}:
\begin{align}\label{sigma-iso}
\hS(\ioj) = 
i \o_{j}\left[1-Z(\ioj) \right] \hat{\tau}_0 + \chi(\ioj) \hat{\tau}_3 + \phi(\ioj) \hat{\tau}_1,
\end{align}
where $Z(\ioj)$ is the mass renormalization function, $\chi(\ioj)$ is the energy shift, and $\phi(\ioj)$ is the order parameter. The quantity $\Delta(\ioj)=\phi(\ioj)/Z(\ioj)$ defines the superconducting gap function. Through the Dyson equation~\eqref{G-1}, the calculation of the Green's function $\hG(\ve,\iojp)$ is reduced to solving three coupled equations for $Z(\ioj)$, $\chi(\ioj)$, and $\phi(\ioj)$. A detailed derivation is given in Sec.~IV of Supplemental Information~\cite{SI}. 
We refer to this final set as the isotropic non-adiabatic full-bandwidth ({\small NA-FBW}) Eliashberg equations and are given by 
\begin{widetext}
\begin{align}
Z(\ioj)
&= 1 + \frac{k_{\rm B}T}{\NF \o_j} \sum_{j'} 
\l(\ioj-\iojp)
\int_{-\infty}^{\infty}\! d\ve' N(\ve') \g^Z(\ve',\iojp)
\nonumber \\ 
& \quad \quad + \frac{(k_{\rm B}T)^2}{\NF^2 \, \o_j}\sum_{j'j''} 
\l^{\rm V}(\ioj -\iojp, \ioj -\iojdp)
\int_{-\infty}^{\infty}\! d\ve' N(\ve') 
\int_{-\infty}^{\infty}\! d(\ve''+\ve'-\ve) N(\ve''+\ve'-\ve)  
\int_{-\infty}^{\infty}\! d\ve''  N(\ve'') 
\nonumber \\ 
& \quad \quad \quad \quad \quad \quad \times
\left [ \g^T(\ve',\iojp) P^Z(\ve''+\ve'-\ve,\iojtp) 
\g(\ve'',\iojdp) \right ],\label{Z-FBW}
\end{align}
\begin{align}
\chi(\ioj) &=
 -\frac{k_{\rm B}T}{\NF} \sum_{j'} 
\l(\ioj-\iojp)
\int_{-\infty}^{\infty}\! d\ve' N(\ve')
\g^\chi(\ve',\iojp)
\nonumber \\ 
& \quad
-\frac{(k_{\rm B}T)^2}{\NF^2} \sum_{j'j''}  
\l^{\rm V}(\ioj -\iojp, \ioj -\iojdp)
\int_{-\infty}^{\infty}\! d\ve' N(\ve') 
\int_{-\infty}^{\infty}\! d(\ve''+\ve'-\ve) N(\ve''+\ve'-\ve)  
\int_{-\infty}^{\infty}\! d\ve''  N(\ve'') 
\nonumber \\ 
& \quad \quad \quad \quad \times
\left [ \g^T(\ve',\iojp) P^\chi(\ve''+\ve'-\ve,\iojtp) 
\g(\ve'',\iojdp) \right ], \label{chi-FBW} 
\end{align}
\begin{align}
\phi(\ioj) 
&= \frac{k_{\rm B}T}{\NF} \sum_{j'} 
\left [ \l(\ioj-\iojp) -\mu_{\rm E}^* \right ]
\int_{-\infty}^{\infty}\! d\ve' N(\ve')
\g^\phi(\ve',\iojp) \nonumber \\
& \hspace{0.5mm} + \frac{(k_{\rm B}T)^2}{\NF^2} \sum_{j'j''}  
\l^{\rm V}(\ioj -\iojp, \ioj -\iojdp) 
\int_{-\infty}^{\infty}\! d\ve' N(\ve') 
\int_{-\infty}^{\infty}\! d(\ve''+\ve'-\ve) N(\ve''+\ve'-\ve)  
\int_{-\infty}^{\infty}\! d\ve''  N(\ve'') 
\nonumber \\ 
& \quad \quad \quad \times
\left [ \g^T(\ve',\iojp) P^\phi(\ve''+\ve'-\ve,\iojtp) 
\g(\ve'',\iojdp) \right ], \label{phi-FBW} 
\end{align}
\begin{align}
N_e &= \int_{-\infty}^{\infty}\! d\ve' N(\ve') 
\left [ 1 - 2 k_{\rm B} T \sum_{j'} \frac{\ve - \mu_{\rm F} + \chi(\iojp)}
{\Theta(\ve, \iojp)} \right]. \label{particle-number}
\end{align}
\end{widetext}
In Eq.~\eqref{particle-number}, $N_e$ represents the electron number that determines the chemical potential which is updated self-consistently~\cite{Marsiglio2008,Lee2023,Lucrezi2024}. This is referred to as {\small FBW}+$\mu$, while calculations performed with a fixed $\mu_{\rm F}= \ef$ are denoted simply as {\small FBW}~\cite{Lee2023,Lucrezi2024}. The Coulomb interaction enters the Eliashberg expression only through Eq.~\eqref{phi-FBW}, where it is approximated by the semi-empirical pseudopotential $\mu_{\rm E}^*$ introduced earlier. The summation over the Matsubara frequencies in Eqs.~\eqref{Z-FBW}$-$\eqref{particle-number} formally extends from $-\infty$ to $+\infty$, however, in practical calculations, it is evaluated over positive Matsubara frequencies only and truncated at a cutoff frequency, as detailed in Sec.~V of the Supplementary Information~\cite{SI}. 
Above, we follow Ref.~\cite{Schrodi2020} and introduce the pseudo-vector $\g(\ve,\ioj)$ as
\begin{align}
\g(\ve,\ioj) &=
\begin{pmatrix}
\g^Z (\ve,\ioj) \\
\g^\chi (\ve,\ioj)\\
\g^\phi (\ve,\ioj)
\end{pmatrix}, 
\end{align}
and its transpose is given by
\begin{align}
\g^T(\ve,\ioj) &=
\begin{pmatrix} 
\g^Z(\ve,\ioj) & \g^\chi (\ve,\ioj) 
& \g^\phi (\ve,\ioj)  
\end{pmatrix}, 
\end{align}
along with the matrices $P^Z(\ve,\ioj)$, $P^\chi(\ve,\ioj)$, and $P^\phi(\ve,\ioj)$ defined as follows
%
\begin{align}
P^Z(\ve,\ioj) &=
\! \begin{pmatrix}
-\g^Z(\ve,\ioj) & \g^\chi(\ve,\ioj) 
& -\g^\phi(\ve,\ioj)  \\
\g^\chi(\ve,\ioj)  & \g^Z(\ve,\ioj) & 0 \\
-\g^\phi(\ve,\ioj) & 0& \g^Z(\ve,\ioj) 
\end{pmatrix}\!, 
\end{align}
\begin{align}
P^\chi(\ve,\ioj) &=
\! \begin{pmatrix}
-\g^\chi(\ve,\ioj)  & -\g^Z(\ve,\ioj)  & 0 \\
-\g^Z(\ve,\ioj) & \g^\chi(\ve,\ioj) & -\g^\phi(\ve,\ioj)  \\
0 & -\g^\phi(\ve,\ioj) & -\g^\chi(\ve,\ioj) 
\end{pmatrix}\! , 
\end{align}
\begin{align}
P^\phi(\ve,\ioj) &=
\!\begin{pmatrix}
\g^\phi(\ve,\ioj) & 0 & -\g^Z(\ve,\ioj)  \\
0& \g^\phi(\ve,\ioj) & \g^\chi(\ve,\ioj) \\
-\g^Z(\ve,\ioj) & \g^\chi(\ve,\ioj) & -\g^\phi(\ve,\ioj) 
\end{pmatrix}. 
\end{align}
%
The elements of $\g(\ve,\ioj)$, $P^Z(\ve,\ioj)$, $P^\chi(\ve,\ioj)$, and $P^\phi(\ve,\ioj)$ are given by
\begin{align}\label{gamma}
\gamma^Z(\ve, \ioj) &= \frac{\o_{j}Z(\ioj)}{\Theta(\ve, \ioj)}, \\
\gamma^\chi(\ve,\ioj) &= \frac{\ve -\mu_{\rm F} + \chi(\ioj)}{\Theta(\ve,\ioj)}, \\
\gamma^\phi(\ve,\ioj) &= \frac{\phi(\ioj)}{\Theta(\ve,\ioj)},
\end{align}
with 
\begin{align}\label{theta-exp}
\Theta(\ve,\ioj) 
&=\left[\o_{j}Z(\ioj)\right]^2 + \left[\ve -\mu_{\rm F} + \chi(\ioj)\right]^2 
+ \left[\phi(\ioj)\right]^2 .
\end{align}

\subsection{Isotropic non-adiabatic Fermi-surface restricted Eliashberg equations}

We can further simplify the isotropic {\small NA-FBW} Eliashberg equations by adopting the constant DOS approximation, such that $N(\ve) \rightarrow \NF$. Assuming an infinite electronic bandwidth at half-filling, the energy integral can be evaluated analytically,  as detailed in Sec.~VI of the Supplementary Information~\cite{SI}, leading to the isotropic non-adiabatic Fermi-surface restricted ({\small NA-FSR}) Eliashberg equations
\begin{align} 
Z(\ioj) = &1 + \frac{\pi k_{\rm B} T}{\o_j} 
\sum_{j'}  \l(\ioj-\iojp)
 \tg^Z(\iojp) \nonumber \\  
&+ \frac{\pi^3(k_{\rm B}T)^2 \NF}{\o_j} 
\sum_{j'} 
\sum_{j''}  
\l^V(\ioj -\iojp, \ioj -\iojdp)   \nonumber \\
&\times \left[ \tg^T(\iojp) \tilde{P}^\o(\iojtp) \tg(\iojdp) \right], 
\label{FSR-iso} 
\end{align}
\begin{align}
\D(\ioj) Z(\ioj)
&=\pi k_{\rm B} T \sum_{j'}
\left [ \l(\ioj-\iojp) -\mu_{\rm E}^* \right ] \tg^\phi(\iojp)
\nonumber \\
&+\pi^3 (k_{\rm B}T)^2 \NF 
\sum_{j'} 
\sum_{j''}  
\l^{\rm V}(\ioj -\iojp, \ioj -\iojdp) \nonumber \\
& \times \left [ \tg^T(\iojp) \tilde{P}^\D(\iojtp) 
\tg(\iojdp) \right ]. \label{delta-iso-FSR}
\end{align}
The pseudo-vector $\tg(\ve,\ioj)$ and its transpose $\tg^T(\ve,\ioj)$, along with the matrices $\tilde{P}^\o(\ioj)$ and $ \tilde{P}^\D(\ioj)$ are expressed as
\begin{align}
\tg(\ioj) =
\begin{pmatrix}
\tg^\o (\ioj) \\
\g^\D (\ioj)
\end{pmatrix}, 
\hspace{0.05cm}
\tg^T(\ioj) =
\begin{pmatrix} 
\g^\o(\ioj) & \g^\D (\ioj) 
\end{pmatrix}, 
\end{align}
\begin{align}\label{Pz-FSR-iso}
    \tilde{P}^\o(\ioj) & =
    \begin{pmatrix}
    -\tg^\o (\ioj) & -\tg^\D (\ioj) \\
    -\tg^\D (\ioj) & \tg^\o (\ioj)
    \end{pmatrix}, 
 \end{align}
 \begin{align}\label{phi-iso-FSR}
    \tilde{P}^\D(\ioj) &=
    \begin{pmatrix}
    \tg^\D (\ioj) & -\tg^\o (\ioj) \\
    -\tg^\o (\ioj) & -\tg^\D (\ioj)
    \end{pmatrix},
\end{align}
where we define
\begin{subequations}
\begin{align}
\tg^\o(\ioj) &= \frac{\o_j}{\sqrt{ \o_j^2 + \D^2(\ioj)}},   \\ 
\tg^\D(\ioj) &= \frac{\D(\ioj)}{\sqrt{\o_j^2 + \D^2(\ioj)}},    
\end{align}
\end{subequations}
with $\Delta(\ioj)=\phi(\ioj)/Z(\ioj)$ being the superconducting gap function.

\section{\label{sec:methods} Computational Methods}

We employed the Quantum {\small ESPRESSO} (QE) package~\cite{Giannozzi2017} with optimized norm-conserving Vanderbilt pseudopotentials ({\small ONCVPSP})~\cite{Hamann2013} from the Hamman's library generated with the Perdew-Burke-Ernzerhof parametrization~\cite{Perdew1996}. We used plane-wave cutoffs of 60 and 100~Ry, a Methfessel-Paxton smearing~\cite{Methfessel1989} value of 0.01~Ry, and $\Gamma$-centered $\bk$-grids of $24 \times 24 \times 24$ and $16 \times 16 \times 16$  to describe the electronic structure of H$_{3}$S and Pb, respectively. The lattice parameters and atomic positions were relaxed until the total energy was converged within $10^{-6}$~Ry and the maximum force on each atom was less than $10^{-4}$ Ry/\AA. The dynamical matrices and the linear variation of the self-consistent potential were calculated within DFPT~\cite{Baroni2001} on $\bq$-meshes of $4 \times 4 \times 4$ and $8 \times 8 \times 8$ for H$_{3}$S and Pb, respectively.
\begin{figure*}[!t]
    \centering 
    \includegraphics[width=0.97\textwidth]{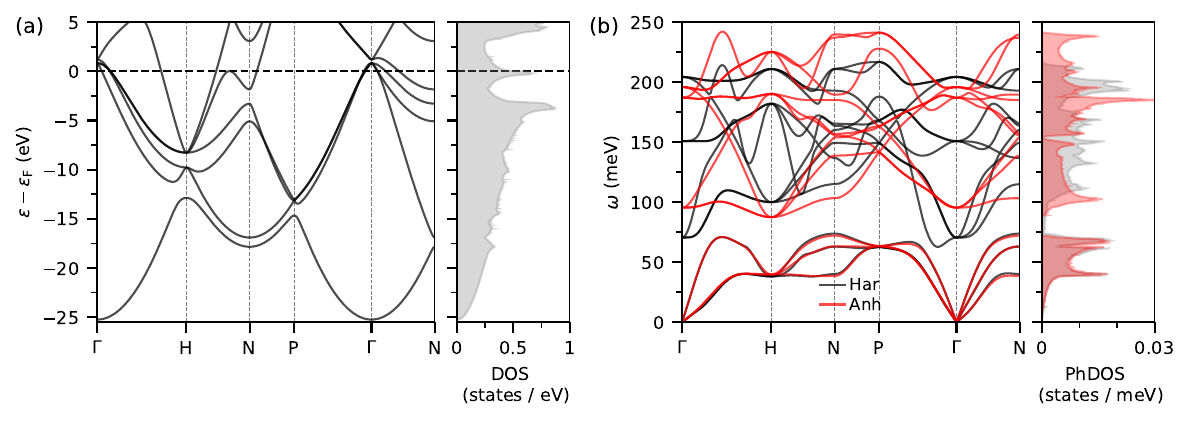}
    \caption{(a) Electronic band structure and total density of states (DOS), highlighting the vHs peak near $\ef$, and 
    (b) phonon dispersion and phonon density of states (PhDOS) computed within the harmonic (black solid lines) and anharmonic approximations (red solid lines) for H$_3$S at 200~GPa. }
    \label{fig:band}
\end{figure*}
For H$_{3}$S at 200~GPa, we carried out a series of convergence tests to ensure the reliability of our phonon calculations. First, we computed the phonon dispersion by varying the $\bq$-mesh from $2 \times 2 \times 2$ to $8 \times 8 \times 8$ (see Fig.~S3(a)~\cite{SI}). Notably, imaginary modes appear when using a $3 \times 3 \times 3$ $\bq$-grid, which are eliminated upon including anharmonic corrections. To reduce the computational cost, we examined the convergence of the phonon dispersion with respect to the kinetic energy cutoff by lowering it from 100~Ry to 60~Ry. The results in Fig.~S3(b)~\cite{SI} show that a 60~Ry cutoff yields converged phonon frequencies and is therefore used throughout this work.
We computed the anharmonic phonons for H$_3$S with the anharmonic special displacement method (A-SDM)~\cite{Zacharias2020,Zacharias2023} implemented in the ZG code of {\textsc EPW}~\cite{Lee2023}. Calculations were performed for $2\times 2 \times 2$, $3\times 3 \times 3$, and $4\times 4 \times 4$ supercells corresponding to 192, 648, and 1,536 distorted configurations, respectively. An $12 \times 12 \times 12$ $\bk$-grid for the primitive unit cell was found to be enough to extract the interatomic force constants (IFCs) for the anharmonic phonon dispersions, drastically reducing the computational cost (see Fig.~S4(a)~\cite{SI}). All anharmonic calculations presented here were performed at 0~K. 
The {\textsc EPW} code~\cite{Giustino2007,Ponce2016,Margine2013,Lee2023} was used to investigate e-ph interactions and superconducting properties. The electronic wavefunctions required for the Wannier interpolation~\cite{Marzari2012,Pizzi2020,Marazzo2024} were obtained on a uniform $\Gamma$-centered $8 \times 8 \times 8$ $\mathbf{k}$-grid. For H$_{3}$S, we used ten atom-centered orbitals to describe the electronic structure, with one $s$ orbital for each H atom and five $s, p, d_{xy}, d_{xz}, d_{yz}$ orbitals for the S atom, while for Pb we used four $sp^3$ orbitals. The isotropic Eliashberg spectral functions were computed using uniform $48 \times 48 \times 48$ $\bk$- and $24 \times 24 \times 24$ $\bq$-point grids for electrons and phonons, respectively. 
The isotropic Eliashberg equations were solved on the imaginary Matsubara frequency axis, using an energy cutoff of 2.5~eV for H$_{3}$S and 1~eV for Pb. Dirac delta functions for electrons and phonons were approximated by Gaussians with widths of 50~meV and 2.5~meV for H$_{3}$S, and 50~meV and 1.5~meV for Pb. In the case of H$_{3}$S, the DOS, e-ph coupling, and superconducting gap are highly sensitive to the position of the Fermi level due to the vHs located near $\ef$. To ensure consistency and numerical accuracy, $\ef$ was set to the value obtained from \textsc{EPW} calculations using a $48 \times 48 \times 48$ $\bk$-grid for both H$_{3}$S and Pb. In addition, we found that to achieve convergence of the superconducting gap and critical temperature in the {\small FBW} calculations, an energy window of at least 1~eV for H$_3$S and Pb is required when computing the DOS.

\section{\label{sec:results}Results and Discussion}

\subsection{\label{sec:band-H3S}Electronic structure and phonon dispersion of \texorpdfstring{H$_3$S}{h3s}}

At 200~GPa, H$_3$S crystallizes in the high-symmetry $Im\bar3m$ phase with a lattice parameter of 3.089 \AA{}, consistent with previous theoretical studies~\cite{Duan2014,Bernstein2015,Errea2015}. 
The band structure, shown in Fig.~\ref{fig:band}(a), exhibits a broad bandwidth of approximately 25~eV below $\ef$, with parabolic dispersion for the deeper lying bands. The corresponding DOS features two prominent valleys at approximately $-$1.75~eV and 2.50~eV, as well as a sharp vHs peak near $\ef$. This vHs significantly enhances the e-ph coupling strength, which is a key factor in the high $\tc$ of H$_3$S~\cite{Sano2016,Quan2016,Akashi2020,Thomsen2024}. At higher pressures, the vHs peak shifts away from $\ef$, leading to a decrease in the DOS at $\ef$ and a corresponding suppression of $\tc$~\cite{Akashi2015}. 
Additionally, the width of the vHs effectively defines a low Fermi energy scale ($\sim$ 300~meV), reducing the Fermi velocity $v_{\rm F}$ and raising questions about the validity of Migdal's approximation in this system~\cite{Jarlborg2016,Gorkov2018}.
Figure~\ref{fig:band}(b) compares the phonon dispersion and phonon density of states (PhDOS), calculated within the harmonic and anharmonic approximations. The phonon spectrum exhibits a clear separation between low-frequency acoustic modes (below 75~meV), mainly arising from sulfur vibrations, and high-frequency optical modes (above 75~meV), primarily involving hydrogen. This separation is a direct consequence of the large mass difference between S and H atoms. 

\begin{figure*}[!t]
    \centering
    \includegraphics[width=0.97\textwidth]{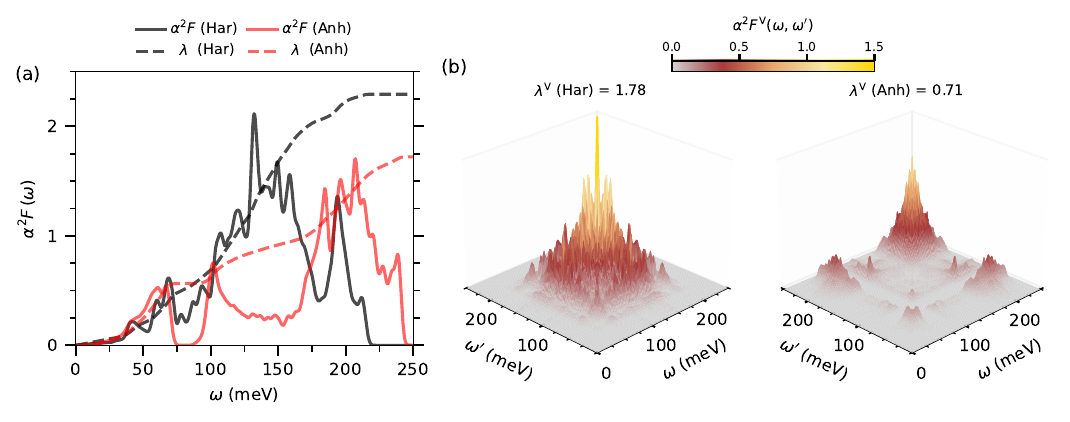}
    \caption{
    (a) Eliashberg spectral function $\a^2F(\o)$ and integrated e-ph coupling strength $\l(\o)$ for both harmonic and anharmonic phonons within the adiabatic Migdal approximation, and 
    (b) vertex-corrected Eliashberg spectral function $\a^2F^{\rm V} (\o, \op)$ and integrated vertex-corrected e-ph coupling strength $\l^{\rm V}$ with harmonic (left) and anharmonic (right) phonons for H$_3$S at 200~GPa.} 
    \label{fig:e-ph}
\end{figure*}
For the anharmonic calculations, convergence was verified with respect to both the atomic displacement amplitude and the supercell size, as detailed in Sec.~VII of the Supporting Information (see  Fig.~S4(b) and ~S5)~\cite{SI}. The low-frequency phonon modes are largely unaffected by anharmonicity, in line with a previous study~\cite{Errea2015}. This insensitivity stems from the heavier mass and smaller vibrational amplitudes of sulfur atoms, which make them less responsive to anharmonic effects. 
In contrast, the high-frequency optical modes undergo substantial anharmonic renormalization, with hardening of up to 30~meV observed along several high-symmetry directions. 
The PhDOS further reinforces this picture. While the spectral features associated with sulfur vibrations remain nearly intact, the high-frequency region shows significant broadening extending up to 243~meV. Accurately capturing these anharmonic effects is essential for a reliable description of vibrational dynamics and superconducting properties of H$_3$S.
\subsection{\label{sec:ep-coupling}Vertex-corrected electron-phonon coupling in \texorpdfstring{H$_3$S}{h3s}}

The Eliashberg spectral function and the corresponding cumulative e-ph coupling strength for H$_3$S at 200~GPa are shown in Fig.~\ref{fig:e-ph}(a). The harmonic $\a^2F(\o)$ features a broad central peak spanning the 75$-$175~meV range, accompanied by a smaller peak at lower frequencies and a distinct sharp peak at 194~meV. As a result, the cumulative $\l(\o)$ increases steadily, with the acoustic region contributing approximately 0.48, and continues to rise throughout the spectrum, ultimately reaching a total value of $\l$ = 2.29, consistent with previous reports~\cite{Duan2014,Flores2016,Sanna2018,Lucrezi2024}. 
When anharmonic effects are included, both phonon frequencies and eigenvectors are renormalized, which in turn modifies the e-ph matrix elements according to Eq.~\eqref{eq:dv}. The cumulative e-ph coupling strength $\l(\o)$ from the sulfur-dominated regions increases slightly to 0.56, consistent with the observation that acoustic phonon frequencies are only weakly affected by anharmonicity. Above 75~meV, $\a^2F(\o)$ exhibits two well-separated peaks, corresponding to the softening of the bond-bending and hardening of bond-stretching H modes, respectively. Among these, the bond-stretching modes contribute most significantly to $\l$. In line with a prior study~\cite{Errea2015}, our anharmonic analysis shows a 25\% reduction in the e-ph coupling strength, resulting in a revised total value of 1.72.
As discussed earlier, the coexistence of high-frequency hydrogen phonon modes and a low Fermi energy raises concerns about the validity of the standard Migdal approximation in H$_3$S~\cite{Sano2016,Jarlborg2016,Durajski2016,Gorkov2018} and requires taking higher-order vertex corrections into account beyond conventional ME theory.
To our knowledge, this work presents the first quantitative computation of the full isotropic lowest-order vertex-corrected Eliashberg spectral function $\a^2F^{\rm V}(\o,\o')$ for this system. 
Figure~\ref{fig:e-ph}(b) shows 3D plots of $\a^2F^{\rm V}(\o,\o')$, comparing results based on harmonic (left panel) and anharmonic (right panel) phonons. In the harmonic case, $\a^2F^{\rm V}(\o,\o')$ displays a broad distribution across the mid- to high-frequency phonon spectrum, with dominant contributions concentrated between 75 and 200~meV. In contrast, the low-frequency acoustic region contributes negligibly, indicating that sulfur modes have minimal impact once vertex corrections are included. The integrated vertex-corrected coupling strength is $\l^{\rm V}$= 1.78, approximately 78\% of the corresponding adiabatic value $\l$ = 2.29.
\begin{figure}[!hbt]
    \centering
    \includegraphics[width=0.46\textwidth]{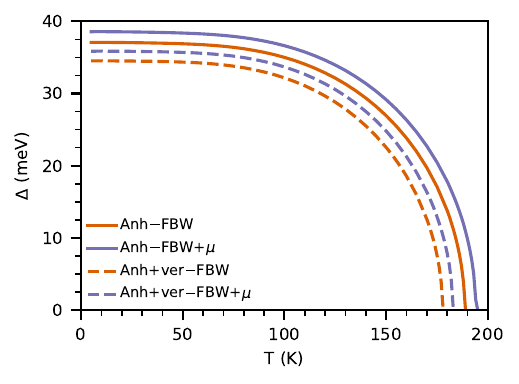}
    \caption{Isotropic superconducting gap $\Delta (T)$ for H$_3$S, calculated with anharmonic phonons using four different approaches: adiabatic {\small FBW} (solid orange), adiabatic {\small FBW}+$\mu$ (solid purple), vertex-corrected {\small FBW} (dashed orange), and vertex-corrected {\small FBW}+$\mu$ (dashed purple) with a Coulomb parameter $\mu_{\rm E}^*$= 0.16. 
    }
    \label{fig:gap}
\end{figure}
In the anharmonic case, $\a^2F^{\rm V}(\o,\o')$ is fragmented, with spectral weight confined in narrower regions of the $(\o,\o')$ plane. Unlike the harmonic scenario, small intensity peaks emerge in the low-frequency acoustic region, reflecting changes in phonon eigenvectors induced by anharmonicity. At mid-frequencies, the spectral weight is reduced and shifts toward higher energies due to the hardening of hydrogen modes. The total vertex-corrected e–ph coupling strength is substantially reduced, with $\l^{\rm V}$= 0.71,  representing nearly a 60\% decrease compared to the harmonic vertex-corrected value.
\begin{table*}[!hbt]
    \caption{Electron-phonon coupling strength ($\l$), vertex-corrected coupling strength ($\l^{\rm V}$), and superconducting $\tc$ computed using adiabatic ({\small FBW}, {\small FBW}+$\mu$, and {\small FSR}), and vertex-corrected (ver-{\small FBW}, ver-{\small FBW}+$\mu$, and ver-{\small FSR}) approaches. For H$_3$S at 200~GPa, we use both harmonic and anharmonic phonons with a Coulomb pseudopotential $\mu_{\rm E}^* =$ 0.16, while for Pb we use harmonic phonons with $\mu_{\rm E}^*=$ 0.10. Experimental $\tc$ values for H$_3$S~\cite{Drozdov2015,Einaga2016,Nakao2019,Du2025} and Pb~\cite{Townsend1962,Khasanov2021} are shown for reference. All the $\tc$ values are in K.}
\label{tab:supercond}
\setlength\tabcolsep{0pt} 
\smallskip 
\begin{tabular*}{\textwidth}{@{\extracolsep{\fill}}l c c c c c  c c c c c c c}
\hline\hline \noalign{\vskip 1mm}
& & & & & \multicolumn{3}{c}{W/o vertex correction} & \multicolumn{3}{c}{With vertex correction} \\
  \cline{6-8} \cline{9-11}
  Material & Phonons & $\l$ & $\l^{\rm V}$ & $\mu^*_{\rm E}$ & $\tc^{\rm {\small FBW}}$ & $\tc^{\rm {\small FBW} +\mu}$ & $\tc^{\rm {\small FSR}}$  & $\tc^{\rm ver-{\small FBW}}$ & $\tc^{\rm ver-{\small FBW}+\mu}$ & $\tc^{\rm ver-{\small FSR}}$ & $\tc^{\rm expt}$ 
  \\ \noalign{\vskip 1mm}
\hline 
\multirow{2}{*}{H$_3$S}
    & Har & 2.29 & 1.78
    & 
    0.16 & 235 & 238 & 246 & 222 & 226 & 198 & \multirow{2}{*}{175-185} \\ 
    & Anh & 1.72 & 0.71 & 0.16 & 189 & 195 & 192 & 178 & 183 & 157 &  \\
    [1mm]
 Pb  & Har & 1.29 & 0.20 & 0.10 & 7.2 & -- & 7.2  & 7.2 & -- &7.2 &7.2 \\ \noalign{\vskip 1mm}  
\hline\hline
\end{tabular*}
\end{table*}
\subsection{Superconducting gap of \texorpdfstring{H$_3$S}{h3s} with vertex corrections}

The superconducting properties of H$_3$S have been extensively investigated, with previous calculations primarily based on the adiabatic ME formalism without vertex corrections, using either the {\small FSR}~\cite{Duan2014,Errea2015} or {\small FBW} approach~\cite{Sano2016,Lucrezi2024,Pellegrini2024}. Although the potential significance of vertex corrections has been acknowledged~\cite{Gorkov2016}, earlier isotropic treatments relied on simplified models, either a single Einstein phonon spectrum~\cite{Sano2016,Ryotaro2017} or empirical modifications to the vertex-corrected e–ph kernel $\lambda^{\rm V}(\ioj - \iojp, \ioj - \iojdp)$~\cite{Durajski2016}. 
In this work, we go beyond such approximations and solve the isotropic Eliashberg equations using the full vertex-corrected e–ph spectral function $\alpha^2F^{\rm V}(\o, \o')$, computed from first principles. 
Several studies on H$_3$S~\cite{Sano2016,Ryotaro2017,Lucrezi2024} have emphasized the importance of the FBW description, owing to its narrow electronic bands and the presence of critical features near $\ef$, such as vHs and Lifshitz transitions~\cite{Gorkov2016,Jarlborg2016}.
To capture these effects, we compute the isotropic superconducting gap $\Delta(T)$ using both the {\small FBW} and {\small FBW}+$\mu$ approaches, with and without vertex corrections, while also incorporating anharmonic phonons, as shown in Fig.~\ref{fig:gap}. These calculations explicitly include the energy dependence of the electronic DOS.
Consistent with previous studies on H$_3$S~\cite{Errea2015,Lucrezi2024}, we adopt a Coulomb pseudopotential value of $\mu^*_{\rm E}$= 0.16. When solving the Eliashberg expressions, this choice, combined with an electronic bandwidth $\ve_{\rm el}$ in the range of 10$-$25~eV corresponds to $\mu_{\rm c}=$ 0.21$-$0.25 via Eqs.~\eqref{eq:muc} and ~\eqref{eq:muE}, in agreement with values computed \textit{ab initio}~\cite{Sanna2018,Pellegrini2023,kogler2025}. 
As shown in Fig.~\ref{fig:gap}, the adiabatic {\small FBW} approach yields a zero-temperature superconducting gap of $\Delta(0)$= 37.1~meV, closing at a critical temperature $\tc \approx$ 189~K. 
When the chemical potential is updated self-consistently within the {\small FBW}+$\mu$ scheme, both the gap and $\tc$ increase slightly to 38.6~meV and 195~K, respectively. The inclusion of vertex corrections 
reduces $\Delta(0)$ to approximately 34.5~meV for the ver-{\small FBW} and 35.9~meV for the ver-{\small FBW}+$\mu$ approach. Consequently, the critical temperature is suppressed by 11$-$12~K relative to the adiabatic case, yielding a $\tc$ of 178~K for the ver-{\small FBW} and 183~K for the ver-{\small FBW}+$\mu$ approach. These theoretical predictions are in excellent agreement with experimental measurements, which report $\tc$ values in the range of 175$-$185~K for H$_3$S under 200$\pm$5~GPa~\cite{Drozdov2015,Nakao2019}.
This level of agreement is particularly significant given that our approach simultaneously accounts for anharmonic phonon effects, non-adiabatic e-ph vertex corrections, and the energy dependence of the DOS. Notably, we find that a one-shot evaluation of the vertex correction substantially underestimates $\tc$, demonstrating that iterative self-consistent calculations are essential for accurately quantifying the impact of e-ph vertex renormalization.
\begin{figure*}[!hbt]
    \centering
    \includegraphics[width=0.9\textwidth]{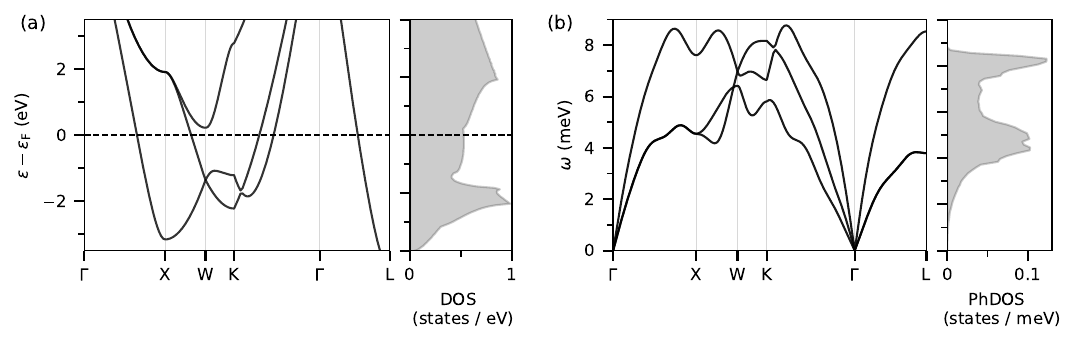}
    \caption{(a) Electronic band structure and total DOS, and (b) phonon dispersion and PhDOS for Pb.}
    \label{fig:pb-band}
\end{figure*}
For comparison, we also analyze the superconducting gap using the constant DOS {\small FSR} approach. As shown in Fig.~S6(a) of the Supporting Information~\cite{SI}, the resulting $\tc$ values lie outside the experimentally observed range~\cite{Drozdov2015}. Without vertex corrections, $\tc$ is slightly above the value obtained with {\small FBW}+$\mu$, while the inclusion of vertex correction leads to a strong underestimation. In Sec.~VIII of the Supporting Information~\cite{SI}, we provide a detailed discussion of the underlying reasons for the limitations of the {\small FSR} approximation. Finally, calculations using harmonic phonons within the {\small FBW}, {\small FBW}+$\mu$, and {\small FSR} approaches, even when incorporating vertex corrections, tend to overestimate the $\tc$, as illustrated in Fig.~S7(b) of the Supporting Information~\cite{SI} and summarized in Table~\ref{tab:supercond}.

Finally, recent experiments on H$_3$S have reported superconducting gap values from tunneling spectra, along with $\tc$ values extracted from electrical resistance measurements~\cite{Du2025}. Since these measurements were performed on samples at 151$-$161~GPa, they are not directly comparable to our calculations at 200~GPa. Nevertheless, the dimensionless BCS ratio 2$\Delta(0)/k_{\rm B} \tc$ provides a meaningful basis for comparison. Using the experimental gap values of 29 and 32~meV with a $\tc =$ 190~K for two samples at 151 and 158~GPa, the corresponding ratios are 3.54 and 3.91, respectively. As summarized in Table~S1 of the Supplemental Information~\cite{SI}, our calculations yield ratios ranging from 4.27 to 4.55 for anharmonic cases and from 4.52 to 5.00 for harmonic ones, consistent with previous Migdal-Eliashberg results at both harmonic and anharmonic levels~\cite{Errea2015,Lucrezi2024}. Further theoretical calculations using different exchange-correlation functionals~\cite{Wang2024}, together with experimental measurements across a broader pressure range, will be essential for resolving the difference in the BCS ratio.

\subsection{Electronic structure and phonon dispersion of Pb}
To further test our hypothesis regarding the breakdown of the Migdal approximation and the significance of e-ph vertex corrections, we analyze elemental Pb, a prototypical conventional superconductor. 
The electronic structure of Pb, shown in Fig.~\ref{fig:pb-band}, features a broad band, with a nearly flat DOS around $\ef$ and sharp features located approximately 2~eV below and above it.
The phonon spectrum extends up to $\sim$9~meV and exhibits a pronounced Kohn anomaly near the W-point in the BZ, which is attributed to strong Fermi-surface nesting~\cite{Aynajian2008}. 
Although Migdal's adiabatic condition ($\hbar \o_0/\ef$) is well satisfied in Pb due to the large separation between the electronic and phononic energy scales, the observed Kohn anomaly suggests the presence of nonlinear e-ph interactions~\cite{Freericks1997}, which are not captured by the standard treatment. These considerations motivate the inclusion of non-adiabatic e-ph vertex effects in our analysis, even for nominally conventional superconductors such as Pb.
\subsection{Electron-phonon coupling and superconductivity with vertex correction for Pb}

\begin{figure*}[!hbt]
    \centering
    \includegraphics[width=\textwidth]{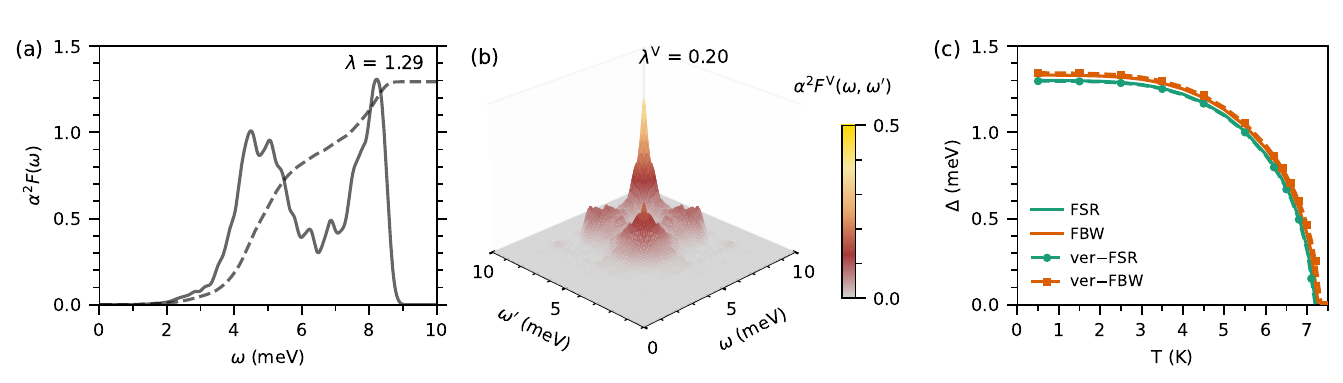}
    \caption{(a) Eliashberg spectral function $\alpha^2F(\o)$ and integrated e-ph coupling constant $\l(\o)$,  
    (b) vertex-corrected Eliashberg spectral function $\a^2F^{\rm V}(\o, \op)$ and integrated vertex-corrected e-ph coupling strength $\l^{\rm V}$, and 
    (c) isotropic superconducting gap $\D(T)$ computed using the adiabatic {\small FSR} (solid teal-green), adiabatic {\small FBW} (solid orange), vertex-corrected {\small FSR} (dashed teal-green with circles), and vertex-corrected {\small FBW} (dashed orange with squares) methods for Pb.}
    \label{fig:pb-eph}
\end{figure*}
To investigate the e-ph interaction in Pb, we computed the isotropic Eliashberg spectral function $\alpha^2F(\o)$ and the cumulative e-ph coupling strength $\l(\o)$. As shown in Fig.~\ref{fig:pb-eph}(a), $\a^2F(\o)$ exhibits a broad peak centered around 4.5~meV (transverse phonons) and a narrower peak at 8~meV (longitudinal phonons), consistent with previous reports~\cite{Liu1996,Heid2010,Margine2013,Mori2024}. The total e-ph coupling strength, $\l$=1.29, indicates strong coupling. 
The high-frequency peak near 7~meV corresponds to the Kohn anomaly at the W point in the phonon dispersion, reflecting enhanced e-ph scattering from Fermi-surface nesting effects~\cite{Aynajian2008}. The cumulative $\l(\o)$ rises sharply below 5~meV, contributing approximately 44\% to the total coupling, confirming the dominant role of the low-frequency phonons in mediating Cooper pairing.
With vertex corrections, the 3D spectral function $\a^2F^{\rm V}(\o, \op)$ (Fig.~\ref{fig:pb-eph}(b)) retains the characteristic peaks at 5 and 8~meV but yields a very low value of $\l^{\rm V}$ = 0.20, representing an 85\% decrease relative to the adiabatic $\l$. This result confirms the validity of Migdal's approximation for Pb, where vertex corrections are negligible. We note that the e-ph coupling in Pb may be slightly enhanced by the inclusion of spin-orbit coupling, as reported in previous studies~\cite{Heid2010}; however, this effect is not expected to alter our conclusions.  
Additionally, the vertex-corrected tunneling inversion study by Freericks et al.~\cite{Freericks1997} found that accounting for the lowest-order vertex correction altered the coupling constant by only $\sim$ 1\% and the $\tc$ by less than 2\%, further reinforcing the robustness of ME theory for conventional superconductors like Pb.
While insightful, their analysis relied on additional simplifying assumptions, such as the constant DOS {\small FSR} approach with the vertex-corrected spectral function taken as the product of adiabatic kernels, which may underestimate the vertex effect.
We performed isotropic Eliashberg calculations to compute the superconducting gap $\Delta(T)$ using a Coulomb parameter $\mu^*_{\rm E}$ = 0.1.
Figure~\ref{fig:pb-eph}(c) shows that the adiabatic {\small FSR} approach without vertex corrections yields a zero-temperature gap $\Delta(0) \approx$ 1.3~meV with a gap ratio of $2\Delta(0)/k_{\rm B}\tc \approx$ 4.3, which exceeds the BCS weak-coupling limit and is in good agreement with the experimental $\tc \approx 7.2$ K~\cite{Townsend1962,Khasanov2021}. Both {\small FSR} and {\small FBW} methods produce nearly similar $\Delta(T)$ curves, reflecting the weak energy dependence of the DOS near the Fermi level and the broad electronic bandwidth of Pb. 
Notably, the inclusion of lowest-order vertex corrections has a negligible impact on the superconducting gap evolution. The vertex-corrected {\small FSR} and {\small FBW} results lie on top of their adiabatic counterparts across all temperatures. This insensitivity of $\Delta(0)$ and $\tc$ to vertex corrections further validates the applicability of ME theory in describing superconductivity in adiabatic and weakly non-adiabatic systems such as Pb.

\section{\label{sec:summary}Summary}
In this work, we revisit the formulation of the lowest-order vertex corrections within the Eliashberg formalism and develop a fully \textit{ab initio} approach to evaluate the vertex-corrected Eliashberg spectral function and the corresponding e-ph coupling strength. This methodology, implemented in the EPW code~\cite{Ponce2016,Lee2023}, enables us to move beyond model Hamiltonians and apply the formalism to realistic materials such as H$_3$S and Pb.
To solve the isotropic Eliashberg equations, we employ three approaches: the {\small FSR}, {\small FBW} with a fixed $\mu_{\rm F}$, and {\small FBW} with a self-consistently updated $\mu_{\rm F}$.
For H$_3$S, calculations include both harmonic and anharmonic phonons using a Coulomb pseudopotential $\mu_{\rm E}^*$= 0.16,  while for Pb, we consider harmonic phonons with $\mu_{\rm E}^*$ = 0.10.
In H$_3$S, which exhibits strong e-ph coupling, incorporating vertex corrections leads to a significant reduction in both $\l$ and $\tc$, signaling a breakdown of the Migdal approximation.
When effects from phonon anharmonicity and variation in the electronic DOS are taken into account, the predicted $\tc$ agrees well with experimental measurements, resolving long-standing discrepancies between theory and experiment.
By contrast, for elemental Pb, a prototypical conventional superconductor, vertex corrections have a negligible impact on both $\l$ and $\tc$, consistent with the expectations of Migdal–Eliashberg theory.
Overall, our findings demonstrate that vertex corrections are essential for accurately describing superconductivity in materials with strong non-adiabaticity, low carrier density, or sharp features in the DOS, such as high-$\tc$ hydrides, while validating their negligible influence in conventional adiabatic superconductors. The computational framework established here provides a rigorous and predictive first-principles methodology for extending the applicability of Eliashberg theory beyond the conventional Migdal limit.

\begin{acknowledgments}
We thank S.~Tiwari and F.~Giustino for insightful discussions on the numerical implementation of the vertex part in the EPW code, R.~Akashi for stimulating conversations and independent verification of our derivations, and M.~Zacharias for assistance with the anharmonic calculations using the ZG~code. S.B.M. acknowledges H.~Paudyal for early discussions on the vertex implementation. This work was primarily supported by the National Science Foundation (NSF) under Award No.~DMR-2035518. Part of this research was supported by the NSF, Office of Advanced Cyberinfrastructure under Grant No. 2103991 of the Cyberinfrastructure for Sustained Scientific (parallelization of the isotropic superconductivity module). Computational resources were provided by the Frontera and Stampede3 supercomputers at the Texas Advanced Computing Center (TACC) at The University of Texas at Austin (http://www.tacc.utexas.edu), supported through the Leadership Resource Allocation (LRAC) award DMR22004 and the ACCESS allocation TG-DMR180071, respectively.
\end{acknowledgments}

\section*{Author Contributions}
R.M. conceived the ideas and secured the funding. R.M. and S.M. derived the theory. S.M. implemented the vertex correction in the EPW code in consultation with H.M. S.M. performed the calculations and drafted the manuscript. R.M. supervised the project and revised the manuscript. All authors were involved in the formal analysis and participated in the review and editing of the final manuscript.


%

\end{document}


\title{Supporting Information: \\Electron-phonon vertex correction effect in superconducting H$_3$S}

\author{Shashi B. Mishra}
\email{mshashi125@gmail.com}
\affiliation{Department of Physics, Applied Physics and Astronomy, Binghamton University-SUNY, Binghamton, New York 13902, USA}
\author{Hitoshi Mori}
\affiliation{Department of Physics, Applied Physics and Astronomy, Binghamton University-SUNY, Binghamton, New York 13902, USA}
\affiliation{Institute for Materials Research, Tohoku University, Sendai 980-8577, Japan}
\author{Elena R. Margine}
\email{rmargine@binghamton.edu}
\affiliation{Department of Physics, Applied Physics and Astronomy, Binghamton University-SUNY, Binghamton, New York 13902, USA}
\date{\today}

\maketitle

\addcontentsline{toc}{section}{}
\tableofcontents
\newpage

\section{\label{sec:diagram}Diagrammatic representation of the electron-phonon self-energy}

\begin{figure}[!ht]
    \centering
    \includegraphics[width=0.8\linewidth]{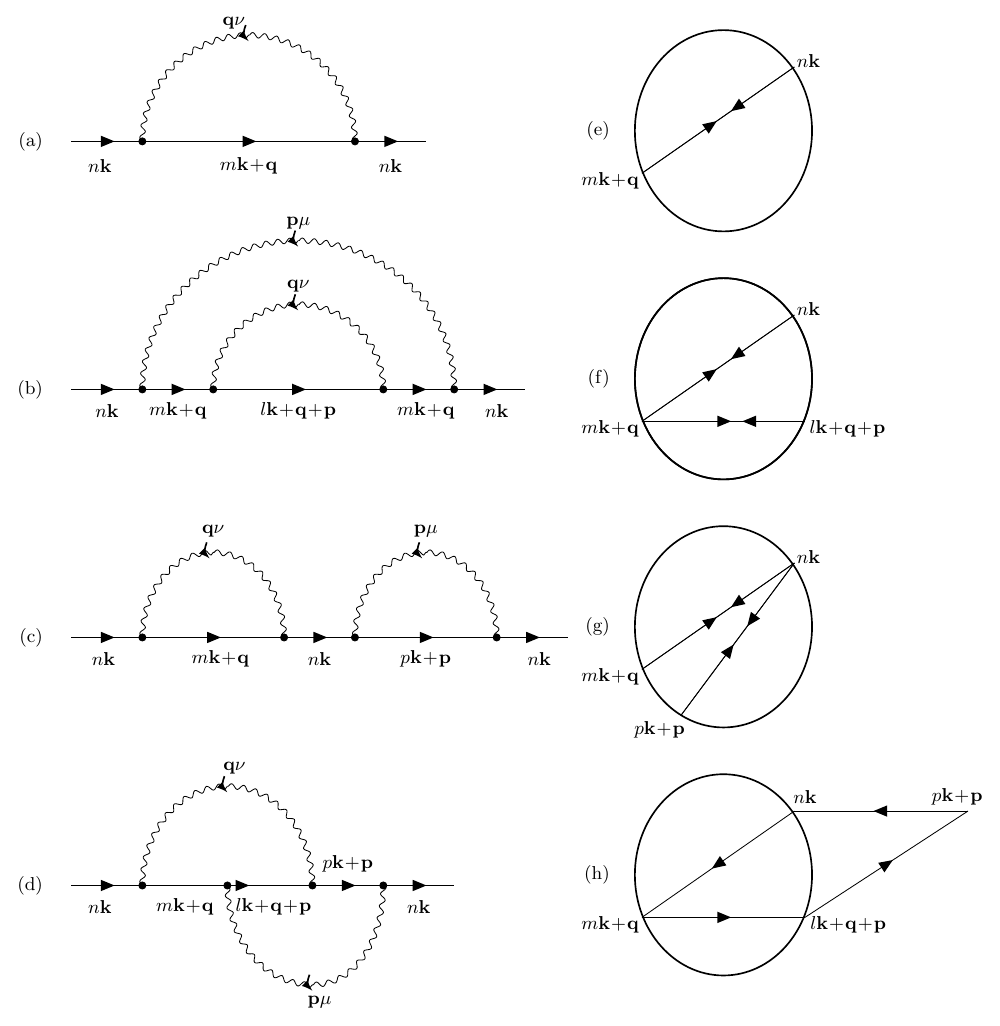}
    \caption{(a-d) Feynman diagrams illustrating the lowest-order electron self-energy contributions, corresponding to Fig.~1(b) in the main text. (e–h) Schematic representations of the associated Fermi surfaces and electronic states involved in the respective scattering processes. Panel (a) depicts the standard Migdal diagram, while panels (b) and (c) show next-order self-energy corrections, whose effects are effectively included in (a). The corresponding scattering processes are shown in (e) and (f), respectively. Panel (d) illustrates the lowest-order vertex correction, with the corresponding scattering representation given in (h).}
    \label{fig:dynson-diagram}
\end{figure}

Figure~\ref{fig:dynson-diagram} presents several lowest-order Feynman diagrams contributing to the electron self-energy, including the standard non-crossing Migdal diagram~\cite{Migdal1958}, and their corresponding scattering processes on the Fermi surface. Diagrams (b) and (c) depict next-order corrections to the Migdal diagram, while (d) represents the lowest-order vertex correction that accounts for scattering between multiple intermediate states beyond the Migdal approximation~\cite{Allen1983}. 
The right panels (e-h) illustrate schematically the associated Fermi surface scattering processes, where arrows indicate the momentum transfer between the electronic states. The diagrams in (f) and (g) demonstrate that these processes are largely confined near the Fermi surface and, as discussed in several Refs.~\cite{Allen1983,Heid2017,Schrodi2020}, lead to small energy shifts that are effectively summed in the Migdal approximation. In contrast, the vertex correction (d) and its associated scattering pathway in (h) involve off-shell intermediate states. These processes, typically neglected in Migdal theory, become increasingly important in systems exhibiting strong non-adiabatic effects or enhanced Fermi-surface nesting.   

\section{\label{sec:iso-der}Derivation of the isotropic self-energy expression}

The self-energy expression given by Eq.~(17) of the main text couples electronic states with momenta $\bk$. To derive the isotropic equations, we simplify the $n\bk$-dependence by defining the energy-resolved average of a quantity $A_{n \bk}$ as~\cite{Pickett1982}
\begin{equation}\label{eq:avg1_form}
    A(\ve) = \frac{ \sum_{n \bk} A_{n \bk} \d(\ve_{n \bk}-\ve)}
    {\sum_{n \bk} \d(\ve_{n \bk}-\ve)} = \frac{ \sum_{n \bk} A_{n \bk} \d(\ve_{n \bk}-\ve)}{N(\ve)}.
\end{equation}
Here, $N(\ve) = \sum_{n \bk} \d(\ve_{n \bk}-\ve)$ denotes the electronic density of states (DOS) per spin at energy $\ve$. 

\noindent Using Eq.~\eqref{eq:avg1_form}, the isotropic form of adiabatic self-energy $\hS_{n \bk}^{\mathrm{(A)}}(\ioj)$ from Eq.~(22) of the main text simplifies to 
\begin{align}\label{self-A-iso}
    \hS^{\mathrm{(A)}}(\ve,\ioj) 
    &= \frac{1}{N(\ve)} \sum_{n \bk} \hS_{n \bk}^{\mathrm{(A)}}(\ioj) \d(\ve_{n \bk}-\ve) 
    \nonumber \\
    &= \frac{k_{\rm B} T}{N(\ve)} 
    \sum_{n m} \sum_{\bk \bq} \sum_{j'} 
    \frac{\l_{n \bk, m \bkq}(\ioj-\iojp)}{\NF} \d(\ve_{n \bk}-\ve)
    \hat{\tau}_3 \hG_{m \bkq}(\iojp) \hat{\tau}_3.
\end{align}
To express this in terms of energy-resolved quantities, we insert the identity $\int_{-\infty}^{+\infty} d\ve' \d(\ve_{m \bkq}-\ve')=1$ into the right-hand side (RHS) of Eq.~\eqref{self-A-iso}, leading to
\begin{align}\label{self-A-iso2}
    \hS^{\mathrm{(A)}}(\ve,\ioj) 
    &= \frac{k_{\rm B} T}{\NF} 
    \int_{-\infty}^{+\infty} d\ve'  \frac{1}{N(\ve)} 
    \sum_{nm } \sum_{\bk \bq} \sum_{j'} 
    \l_{n \bk, m \bkq}(\ioj-\iojp)
    \d(\ve_{n \bk}-\ve) 
    \d(\ve_{m \bkq}-\ve')
    \hat{\tau}_3 \hG_{m \bkq}(\iojp) \hat{\tau}_3
    \nonumber \\
    &= \frac{k_{\rm B} T}{\NF} 
    \int_{-\infty}^{+\infty}  d\ve' 
    \frac{N(\ve')}{N(\ve) N(\ve')} 
    \sum_{nm } \sum_{\bk \bq} \sum_{j'} 
    \l_{n \bk, m \bkq}(\ioj-\iojp)
    \d(\ve_{n \bk}-\ve)
    \d(\ve_{m \bkq}-\ve') 
    \hat{\tau}_3 \hG_{m \bkq}(\iojp) \hat{\tau}_3.
\end{align}
In the second line of Eq.~\eqref{self-A-iso2}, we have inserted the factor $\frac{N(\ve')}{ N(\ve')}$ to facilitate simplification. To proceed, we define a double average over $\bk$ and $\bkq$ on the RHS of Eq.~\eqref{self-A-iso2} as follows
\begin{align} \label{adia-iso-approx}
     &\frac{1}{N(\ve) N(\ve')} 
     \sum_{nm } 
     \sum_{\bk \bq} 
     \l_{n \bk, m \bkq}(\ioj-\iojp)
    \hG_{m \bkq}(\iojp) 
    \d(\ve_{n \bk}-\ve) 
    \d(\ve_{m \bkq}-\ve') 
    \approx  \l(\ve,\ve',\ioj-\iojp) \hG(\ve',\iojp).
\end{align}
This approximation allows us to decouple the integration over the electron-phonon (e-ph) kernel and the electron Green's function in the energy domain. 
Following Allen and Mitrović~\cite{Allen1983}, we neglect the energy dependence of the e-ph kernel $\l(\ve,\ve',\ioj-\iojp)$ and evaluate it at the Fermi level: 
\begin{align}\label{iso-l-approx-ef}
    \l(\ve,\ve',\ioj-\iojp) \approx \l(\ef,\ef,\ioj-\iojp) 
    \equiv \l(\ioj-\iojp).
\end{align}
This approximation is justified by the assumption that only electronic states near Fermi level ($\ef$) contribute significantly to the e-ph interaction; that is, $\d(\ve_{n \bk}-\ve)\rightarrow \d(\ve_{n \bk}-\ef)$. 
Furthermore, the $\bk$-dependence of the Green's function $\hat{G}_{n \bk}(\ioj)$ (see Eq.~(11) in the main text) enters through the band energy $\ve_{n \bk}$. Thus, the dominant $\ve'$ dependence of the energy-resolved Green's function $\hG(\ve',\iojp)$ arises from the explicit appearance of $\ve'$. 
\par\vspace{1em}
\noindent Substituting these approximations into Eq.~\eqref{self-A-iso2}, we obtain the isotropic adiabatic self-energy:
\begin{align}\label{self-A-iso-simplfied} 
    \hS^{\mathrm{(A)}}(\ioj) 
    =& \frac{k_{\rm B} T}{\NF} \sum_{j'} \l(\ioj-\iojp)
    \int_{-\infty}^{+\infty} d\ve' 
    N(\ve')  \hat{\tau}_3 \hG(\ve',\iojp) \hat{\tau}_3.
\end{align}
Here, the isotropic e-ph kernel $\l(\ioj-\iojp)$ is defined as 
\begin{align}\label{l-A-iso}
    \l(\ioj-\iojp) 
    &= \frac{\sum_{nm } \sum_{\bk \bq} 
    \d(\ve_{n \bk}-\ef) \d(\ve_{m \bkq}-\ef)
    \l_{n \bk, m \bkq}(\ioj-\iojp)}
    {\sum_{nm } \sum_{\bk \bq}
    \d(\ve_{n \bk}-\ef) \d(\ve_{m \bkq}-\ef)} \nonumber \\
    &= \int_{0}^{\infty} 
    d\o \frac{2\o}{(\o_j - \o_j')^2 + \o^2}
    \, \a^2F(\o),
\end{align}
where $\a^2F(\o)$ is the isotropic (double-averaged) Eliashberg spectral function expressed as follows:
\begin{align}\label{a2f-A-iso}
    \a^2F(\o) & = 
 \frac{\sum_{nm } \sum_{\bk \bq}
  \d(\ve_{n \bk}-\ef)
  \d(\ve_{m \bkq}-\ef)
  \a^2F_{n \bk, m \bkq}(\o)}{{ \sum_{nm } \sum_{\bk \bq}
    \d(\ve_{n \bk}-\ef) \d(\ve_{m \bkq}-\ef)}} \nonumber \\
  &= \frac{1}{\NF}
\sum_{nm } \sum_{\bk \bq \nu}
  \d(\ve_{n \bk}-\ef)
  \d(\ve_{m \bkq}-\ef)
  |g_{m\bkq,n \bk}^{\bq \nu}|^2 \d(\o-\o_{\bq \nu }).
\end{align}

\par\vspace{1em}

\noindent We now follow similar steps to define the isotropic average of the non-adiabatic self-energy $\hS_{n \bk}^{\mathrm{(NA)}}(\ioj)$, as given in Eq.~(26) of the main text:
\begin{align}\label{self-NA-iso}
\hS^{\mathrm{(NA)}}(\ve, \ioj) 
&= \frac{ \sum_{n \bk} 
\hS_{n \bk}^{\mathrm{(NA)}}(\ioj) \d(\ve_{n \bk}-\ve)}
{ \sum_{n \bk} \d(\ve_{n \bk}-\ve) }
= \frac{\sum_{n \bk} 
\hS_{n \bk}^{\mathrm{(NA)}}(\ioj) \d(\ve_{n \bk}-\ve)}
{N(\ve) }\nonumber \\
&= \frac{(k_{\rm B} T)^2}{N(\ve)} 
\sum_{n \bk} \d(\ve_{n \bk}-\ve)
\sum_{mlp}\sum_{\bq \bp} 
 \sum_{j'j''} 
\frac{\l^{\rm V}_{n\bk, m\bkq, l\bkqp, p\bkp}(\ioj -\iojp, \ioj -\iojdp)}{\NF^2} 
\nonumber \\
& \quad \quad \quad \quad \quad \quad \quad \times 
\hat{\tau}_3\hG_{m \bkq}(\iojp)\hat{\tau}_3  
\hG_{l\bkqp}(\iojtp) 
\hat{\tau}_3\hG_{p \bkp}(\iojdp)\hat{\tau}_3.
\end{align}

\noindent In the second step, we insert the identities $\int_{-\infty}^{+\infty} d\ve' \d(\ve_{m \bkq}-\ve')=1$, 
$\int_{-\infty}^{+\infty} d(\etp) \d(\ve_{l \bkqp}-(\etp))=1$, and $\int_{-\infty}^{+\infty} d\ve'' \d(\ve_{p \bkp}-\ve'')=1$ into the RHS of Eq.~\eqref{self-NA-iso} to get
\begin{align}\label{self-NA-iso-simplify}
    \hS^{\mathrm{(NA)}}(\ve, \ioj) 
    &= \frac{(k_{\rm B} T)^2}{N(\ve) \NF^2} 
    \sum_{nmlp}\sum_{\bk \bq \bp}\sum_{j'j''} 
    \d(\ve_{n \bk}-\ve)
    \int_{-\infty}^{+\infty} d\ve' \d(\ve_{m \bkq}-\ve')
    \int_{-\infty}^{+\infty} d\ve'' \d(\ve_{p \bkp}-\ve'') \nonumber \\
    &\quad \times 
    \int_{-\infty}^{+\infty} d(\etp )
    \d(\ve_{l \bkqp}-(\etp) )   
     \l^{\rm V}_{n\bk, m\bkq, l\bkqp, p\bkp}
     (\ioj -\iojp, \ioj -\iojdp)  
    \nonumber \\
    &  \quad \times \hat{\tau}_3\hG_{m \bkq}(\iojp)\hat{\tau}_3  
\hG_{l\bkqp}(\iojtp) 
\hat{\tau}_3\hG_{p \bkp}(\iojdp)\hat{\tau}_3
    \nonumber \\
    &= \frac{(k_{\rm B} T)^2}{\NF^2}  
    \sum_{j'j''} 
    \int_{-\infty}^{+\infty} d\ve' 
    \int_{-\infty}^{+\infty} d\ve''
    \int_{-\infty}^{+\infty} d(\etp)
    \frac{N(\ve') N(\ve'') N(\etp)}{N(\ve)N(\ve') N(\ve'') N(\etp)} 
    \nonumber \\
    &\quad \times  
    \sum_{nmlp}
    \sum_{\bk \bq \bp}\sum_{j'j''} 
     \d(\ve_{n \bk}-\ve)
     \d(\ve_{m \bkq}-\ve')
    \d(\ve_{p \bkp}-\ve'') 
    \d(\ve_{l \bkqp}-(\etp))    
    \nonumber \\
    &\quad \times  
     \l^{\rm V}_{n\bk, m\bkq, l\bkqp, p\bkp}(\ioj -\iojp, \ioj -\iojdp)
    \nonumber \\
    &  \quad \times 
    \hat{\tau}_3\hG_{m \bkq}(\iojp)\hat{\tau}_3  
\hG_{l\bkqp}(\iojtp) 
\hat{\tau}_3\hG_{p \bkp}(\iojdp)\hat{\tau}_3,
\end{align}
where we introduced $\frac{N(\ve')N(\ve'') N(\etp)}{ N(\ve')N(\ve'') N(\etp)}$ for simplification. As in the adiabatic case, we decouple the integration over the e-ph kernel and the electron Green's function to further simplify the expression. For this purpose, we define the energy-averaged over the RHS of Eq.~\eqref{self-NA-iso-simplify} as follows
\begin{align}\label{NA-iso-approx3}
&\frac{1}{N(\ve) N(\ve') N(\ve'') N(\etp)}
\sum_{nmlp}\sum_{\bk \bq \bp}\sum_{j'j''} 
\l^{\rm V}_{n\bk, m\bkq, l\bkqp, p\bkp}(\ioj -\iojp, \ioj -\iojdp)
\nonumber \\
&\quad \quad  \times 
\d(\ve_{n \bk}-\ve)
     \d(\ve_{m \bkq}-\ve')
    \d(\ve_{p \bkp}-\ve'') 
    \d(\ve_{l \bkqp}-(\etp))  
\nonumber \\
&\quad \quad \times
     \hat{\tau}_3\hG_{m \bkq}(\iojp)\hat{\tau}_3  
\hG_{l\bkqp}(\iojtp) 
\hat{\tau}_3\hG_{p \bkp}(\iojdp)\hat{\tau}_3  
    \nonumber \\ 
&\approx \l^{\rm V}(\ve,\ve',\ve'',\etp,\ioj -\iojp, \ioj -\iojdp)
\hG(\ve',\iojp) \hG(\etp,\iojtp) \hG(\ve'',\iojdp). 
\end{align}
\par\vspace{1em}
\noindent 
As before, we assume that only electronic states close to the Fermi surface contribute to the integral in Eq.~\eqref{NA-iso-approx3} (i.e., $\d(\ve_{n \bk}-\ve)\rightarrow \d(\ve_{n \bk}-\ef)$, 
$\d(\ve_{m \bkq}-\ve')\rightarrow \d(\ve_{m \bkq}-\ef)$,
$\d(\ve_{p \bkp}-\ve'')\rightarrow \d(\ve_{p \bkp}-\ef)$, and 
$\d(\ve_{l \bkqp}-\ve''')\rightarrow \d(\ve_{l \bkqp}-\ef)$) and neglect the energy dependence of the vertex-corrected e-ph kernel $\l^{\rm V}$. Thus,
\begin{align}\label{iso-l-NA-approx-ef}
    \l^{\rm V}(\ve,\ve',\ve'',\etp,\ioj -\iojp, \ioj -\iojdp) &\approx 
    \l^{\rm V}(\ef,\ef,\ef,\ef,\ioj -\iojp, \ioj -\iojdp) \nonumber \\
    &\equiv \l^V(\ioj-\iojp, \ioj -\iojdp).
\end{align}

\noindent
Furthermore, as discussed in the adiabatic case, the Green's function depends on $n\bk$ only through the band energy $\ve_{n\bk}$ (see Eq.~(11) in the main text), so the dominant $\ve$ dependence of energy-resolved Green's function $\hG(\ve,\ioj)$ arises from the explicit appearance of $\ve$ (see Sec.~\ref{sec:eliash-soln}). 
\par\vspace{1em}
\noindent Therefore, the isotropic non-adiabatic self-energy Eq.~\eqref{self-NA-iso-simplify} simplifies to 
\begin{align} \label{self-NA-iso-simplified}
\hS^{\mathrm{(NA)}}(\ioj) =  
\frac{(k_{\rm B} T)^2}{\NF^2} & \sum_{j'j''} 
\l^{\rm V}(\ioj -\iojp, \ioj -\iojdp)
\int_{-\infty}^{+\infty} d\ve' N(\ve') 
\int_{-\infty}^{+\infty} d(\etp) N(\etp)
\int_{-\infty}^{+\infty} d\ve'' N(\ve'')
\nonumber \\
& \times 
\hat{\tau}_3\hG(\ve',\iojp)\hat{\tau}_3\hG(\etp,\iojtp) \hat{\tau}_3\hG(\ve'',\iojdp)\hat{\tau}_3,
\end{align}
where the isotropic vertex corrections to the e-ph kernel $\l^{\rm V}(\ioj-\iojp, \ioj -\iojdp)$ is given by Eq.~(32) of the main text.  

\section{\label{sec:sym-a2fv}Symmetry reduction of the vertex-corrected Eliashberg spectral function}

The isotropic vertex-corrected Eliashberg spectral function $\a^2F^{\rm V}(\o, \o')$ is given in Eq.~(34) of the main text.
We exploit the invariance under the exchange $(\bq, \nu, \o) \leftrightarrow (\bp, \mu, \op)$, which can be understood by examining the corresponding Feynman diagram. The vertex-corrected process, shown in Fig. 1(b) of the main text, involves four electron lines connected by two phonon propagators $D^0_{\bq\nu}$ and $D^0_{\bp\mu}$. Exchanging the roles of these two phonons in the diagram yields a topologically equivalent physical process, as the electron loop remains closed and the same four electronic states participate in the scattering. This symmetry transformation does not result in a loss of generality due to the complete summation of all phonon modes and electronic states. Under this exchange, the spectral function can be written as:
\begin{align}\label{eq:symmetry-expanded}
\a^2F^{\rm V}(\o, \op) &= 
\frac{1}{2} \times \frac{1}{C(0)}
  \sum_{n mlp} \sum_{\bk \bq \bp} 
  \d(\ve_{n \bk}-\ef) 
  \d(\ve_{m \bkq}-\ef)
 \d(\ve_{l \bkqp}-\ef) 
 \d(\ve_{p \bkp}-\ef)  
  \nonumber \\
 &\quad \quad \times 
 \NF^2 \sum_{\nu\mu}
 g_{m\bkq,n\bk}^{\bq \nu}
 g_{l\bkqp, m\bkq}^{\bp \mu}
g_{p\bkp, l\bkqp}^{-\bq \nu}
g_{n\bk, p\bkp}^{-\bp \mu}  
\delta(\o-\o_{\bq \nu}) \delta(\op-\o_{\bp \mu})
\nonumber \\
& + 
\frac{1}{2} \times \frac{1}{C(0)}
    \sum_{n mlp} \sum_{\bk \bp \bq } 
  \d(\ve_{n \bk}-\ef) 
  \d(\ve_{m \bk+\bp}-\ef)
 \d(\ve_{l \bk+\bp+\bq}-\ef) 
 \d(\ve_{p \bk+\bq}-\ef)     
  \nonumber \\
 &\quad \quad \times 
 \NF^2 \sum_{\nu\mu}
    g_{m\bkp,n\bk}^{\bp \mu}
g_{l\bk+\bp+\bq, m\bkp}^{\bq \nu}
g_{p\bkq, l\bk+\bp+\bq}^{-\bp \mu}
g_{n\bk, p\bkq}^{-\bq \nu}  
\d(\op-\o_{\bp \mu}) \d(\o-\o_{\bq \nu}).
\end{align}

\noindent Applying the Hermitian property of the e-ph matrix elements, $g_{m\bkq, n\bk}^{\bq\nu} = \left[g_{n\bk, m\bkq}^{-\bq\nu}\right]^*$, the second term becomes:
\begin{align}\label{eq:hermitian-applied}
&\frac{1}{2} \times \frac{1}{C(0)}
    \sum_{n mlp} \sum_{\bk \bp \bq } 
  \d(\ve_{n \bk}-\ef) 
  \d(\ve_{m \bk+\bp}-\ef)
 \d(\ve_{l \bk+\bp+\bq}-\ef) 
 \d(\ve_{p \bk+\bq}-\ef) 
  \nonumber \\
 &\quad \times 
 N_F^2 \sum_{\nu\mu}
    \left[g_{n\bk, m\bk+\bp}^{-\bp \mu}\right]^*
\left[g_{m\bk+\bp, l\bk+\bp+\bq}^{-\bq \nu}\right]^*
\left[g_{l\bk+\bp+\bq, p\bk+\bq}^{\bp \mu}\right]^*
\left[g_{p\bk+\bq, n\bk}^{\bq \nu}\right]^*  
\delta(\op-\o_{\bp \mu}) \delta(\o-\o_{\bq \nu}).
\end{align}

\noindent Since the vertex-corrected spectral function $\alpha^2F^{\rm V}(\omega, \omega')$ must be real-valued as it represents a physical observable, we can combine the original term and its complex conjugate to obtain:
\begin{align}\label{eq:final-result}
\alpha^2F^{\rm V}(\o, \op) &= 
\frac{N_F^2}{C(0)}
  \sum_{n mlp} \sum_{\bk \bq \bp} 
  \d(\ve_{n \bk}-\ef) 
  \d(\ve_{m \bkq}-\ef)
 \d(\ve_{l \bkqp}-\ef) 
 \d(\ve_{p \bkp}-\ef)   
  \nonumber \\
 &\quad \times 
\text{Re}\left[
g_{m\bkq,n\bk}^{\bq \nu}
g_{l\bkqp, m\bkq}^{\bp \mu}
g_{p\bkp, l\bkqp}^{-\bq \nu}
g_{n\bk, p\bkp}^{-\bp \mu}  \right]
\d(\o-\o_{\bq \nu}) \d(\op-\o_{\bp \mu}).
\end{align}
This symmetry-reduced expression allows us to disregard the phase factor in the product of e-ph matrix elements, thereby reducing the computational cost by summing only over identical triplets $(\mathbf{q}, \nu, \omega)$ and $(\mathbf{p}, \mu, \omega')$, and eliminating duplicate contributions without loss of physical content. 

\section{\label{sec:eliash-soln}Simplification of the vertex-corrected self-energy within Eliashberg formalism} 

The interacting Green's function $\hG_{n\bk}(\ioj)$, given in Eq.~(11) of the main text, can be written under the isotropic approximation as
\begin{equation}\label{G-1}
\left[\hG(\ve,\ioj)\right]^{-1}  
= \left[\hG^{0}(\ve,\ioj)\right]^{-1} - \hS(\ioj),
\end{equation}
where the isotropic non-interacting Green's function $\hG^{0}(\ve, \ioj)$ reduces to
\begin{equation}\label{G-0}
\left[\hG^{0}(\ve, \ioj)\right]^{-1} = i\o_{j}\hat{\tau}_0 - (\ve - \mu_{\rm F})\hat{\tau}_3.
\end{equation}

\noindent Substituting Eq.~\eqref{G-0} and the self-energy expression from Eq.~(36) of the main text into Eq.~\eqref{G-1}, we obtain the Dyson equation 
\begin{equation}\label{G-1-b}
\left[\hG(\ve, \ioj)\right]^{-1}  
= i\o_{j}Z(\ioj)\hat{\tau}_0 
- \left[\ve - \mu_{\rm F} + \chi(\ioj)\right]\hat{\tau}_3 
- \phi(\ioj) \hat{\tau}_1.
\end{equation}
\noindent Inverting this $2 \times 2$ matrix, we get
\begin{align}\label{Gnk}
\hG(\ve,\ioj) &= 
\frac{i\o_{j}Z(\ioj) \hat{\tau}_0 
+\left[\ve - \mu_{\rm F} + \chi(\ioj) \right] \hat{\tau}_3 
+ \phi(\ioj) \hat{\tau}_1 }{\det \left[\hG(\ve,\ioj)\right]^{-1}}
= \frac{-i\o_{j}Z(\ioj) \hat{\tau}_0 
-\left[\ve - \mu_{\rm F} + \chi(\ioj) \right] \hat{\tau}_3 
- \phi(\ioj) \hat{\tau}_1}
{\Theta(\ve,\ioj)},
\end{align}
where 
\begin{align}\label{theta-exp}
\Theta(\ve,\ioj) &= 
-\det\left[\hG(\ve,\ioj)\right]^{-1} 
%
=\left[\o_{j}Z(\ioj)\right]^2 
+ \left[\ve -\mu_{\rm F} + \chi(\ioj)\right]^2 
+ \left[\phi(\ioj)\right]^2 
\end{align}
Using the definitions of $\g^Z(\ve, \ioj)$, $\g^\chi(\ve,\ioj)$, and $\g^\phi(\ve,\ioj)$, as given in Eqs.~(46)$-$(48) of the main text, the Green's function in Eq.~\eqref{Gnk} reduces to
\begin{align}\label{G-final}
\hG(\ve,\ioj) = 
-\left[i\g^Z (\ve, \ioj) \hat{\tau}_0 
+ \g^\chi(\ve, \ioj) \hat{\tau}_3 
+ \g^\phi(\ve,\ioj) \hat{\tau}_1 \right].
\end{align}
\noindent To simplify the product $\hat{\tau}_3 {\hG}(\ve,\ioj) \hat{\tau}_3$ in Eq.~\eqref{G-final}, we use
\begin{align}\label{Gtau}
\hat{\tau}_3 {\hG}(\ve,\ioj) \hat{\tau}_3
&= -\hat{\tau}_3 \left[i\g^Z(\ve,\ioj) \hat{\tau}_0 
+ \g^\chi(\ve,\ioj) \hat{\tau}_3 
+ \g^\phi(\ve,\ioj) \hat{\tau}_1\right] \hat{\tau}_3 
= (-1) \left[ i\g^Z(\ve, \ioj)\hat{\tau}_0
+\g^\chi(\ve, \ioj)\hat{\tau}_3
-\g^\phi(\ve, \ioj)\hat{\tau}_1 \right].
\end{align}

\begin{figure}[!hbt]
\centering
\includegraphics[width=0.16\textwidth]{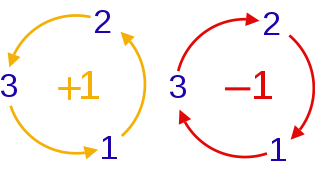}
\caption{\label{fig-2}$\epsilon_{ijk}$ is $+1$ for cyclic permutations of $(1, 2, 3)$ (left), $-1$ for anti-cyclic permutations (right) and $0$ otherwise.}
\end{figure}

\noindent The Pauli matrices obey the identity,
\begin{align}
\hat{\tau}_i \hat{\tau}_j = i\epsilon_{ijk}\hat{\tau}_k + \delta_{ij}\hat{\tau}_0,
\end{align}
where $\epsilon_{ijk}$ is the Levi-Civita symbol, as illustrated in Fig.~\ref{fig-2}. 
\par\vspace{1em}
Using these algebraic relations, we simplify the triple product of Green's functions in Eq.~(31) of the main text: 
\begin{align} \label{G-simpl}
-\hat{\tau}_3 & \hat{G}(\ve', \ioj) \hat{\tau}_3
\hat{G}(\etp, \iojtp) \hat{\tau}_3
\hat{G}(\ve'', \iojdp) \hat{\tau}_3
= a_0 \hat{\tau}_0 + a_1 \hat{\tau}_1 + a_2 \hat{\tau}_3,
\end{align}
where the coefficients $a_0$, $a_1$, and $a_2$ are given by
\begin{subequations}
\begin{align}
a_0 =&  -i\g_{1}^Z \g_{2}^Z    \g_{3}^Z 
      +i\g_{1}^Z \g_{2}^\chi \g_{3}^\chi
      -i\g_{1}^Z \g_{2}^\phi \g_{3}^\phi
      +i \g_{1}^\chi \g_{2}^Z   \g_{3}^\chi 
      +i \g_{1}^\chi \g_{2}^\chi \g_{3}^Z
      +i\g_{1}^\phi \g_{2}^Z \g_{3}^\phi
      -i\g_{1}^\phi \g_{2}^\phi \g_{3}^Z,    \label{tau_0} \\
 a_1 = & \g_{1}^Z \g_{2}^Z \g_{3}^\phi 
    -\g_{1}^Z \g_{2}^\phi \g_{3}^Z 
    -\g_{1}^\chi \g_{2}^\chi \g_{3}^\phi
    -\g_{1}^\chi \g_{2}^\phi \g_{3}^\chi
    +\g_{1}^\phi \g_{2}^Z \g_{3}^Z
    -\g_{1}^\phi \g_{2}^\chi \g_{3}^\chi
    +\g_{1}^\phi \g_{2}^\phi \g_{3}^\phi, \label{tau_1} \\
a_2 = & - \g_{1}^Z \g_{2}^Z \g_{3}^\chi 
   -\g_{1}^Z \g_{2}^\chi \g_{3}^Z
   -\g_{1}^\chi \g_{2}^Z   \g_{3}^Z 
   +\g_{1}^\chi \g_{2}^\chi\g_{3}^\chi
   -\g_{1}^\chi \g_{2}^\phi \g_{3}^\phi
   -\g_{1}^\phi \g_{2}^\chi\g_{3}^\phi
   -\g_{1}^\phi \g_{2}^\phi \g_{3}^\chi. \label{tau_3}
\end{align}
\end{subequations}
Here, for brevity, we define 
$\g_{1}^{Z/\chi/\phi} \equiv \g_{1}^{Z/\chi/\phi} (\ve', \iojp)$, 
$\g_{2}^{Z/\chi/\phi} \equiv \g_{2}^{Z/\chi/\phi} (\etp, \iojtp)$, 
and $\g_{3}^{Z/\chi/\phi} \equiv \g_{3}^{Z/\chi/\phi} (\ve'', \iojdp)$. 
Finally, we note that the $\hat{\tau}_2$ component is neglected because the functions $\phi(\ve, \ioj)$ and $\bar{\phi}(\ve, \ioj)$ are proportional within an arbitrary phase. Without loss of generality, we set the relative phase such that $\bar{\phi}(\ve, \ioj) = 0$~\cite{Allen1983,Ponce2016,Schrodi2020,Lucrezi2024}. 

Following the standard Eliashberg procedure, we equate the coefficients of $\hat{\tau}_0$, $\hat{\tau}_3$, and $\hat{\tau}_1$ in the self-energy expressions (Eqs.~\eqref{self-A-iso-simplfied} and ~\eqref{self-NA-iso-simplified}) to obtain the self-consistent Eliashberg equations for $Z(\ioj)$, $\chi(\ioj)$, and $\phi(\ioj)$, which are given in Eqs.~(37)-(39) of the main text. 
%

\section{\label{sec:matsu}Summation procedure over Matsubara frequencies}

In the imaginary-time formalism of finite-temperature quantum field theory, Matsubara frequencies arise from the Fourier transformation of time-dependent correlation functions. For fermions, the Matsubara frequencies are given by
\begin{equation}\label{eq:mats}
    \ioj = i(2j + 1) \pi k_{\rm B} T 
\end{equation}
where $j$ is an integer and $T$ is the temperature. For bosons, the corresponding Matsubara frequencies are even as given by the formula
\begin{equation}\label{mats-boson}
    i\ot_l = i(2l) \pi k_{\rm B} T 
\end{equation}

The self-energy expressions involve summations over fermionic and bosonic Matsubara frequencies associated with electronic and phononic Green’s functions, respectively. In practical calculations, these infinite sums are truncated by introducing a cutoff frequency $\o_{\rm c}$, typically chosen to be ten times the maximum phonon frequency to ensure convergence of the Eliashberg equations.

The fermionic Matsubara summation is performed over the range $j \in [-N_c - 1, N_c]$, where the cutoff index $N_c$ is defined as
\begin{align}
    \o_{j_{max}} \leq \o_{\rm c} \quad \quad
    &\implies (2j_{max}+1)\pi k_{\rm B} T \leq \o_{\rm c} \nonumber \\
    &\implies j_{max} = \textrm{INT}\left[\frac{1}{2}\left(\frac{\o_{\rm c}}{\pi k_{\rm B}T} - 1\right) \right] \equiv N_c
\end{align}

Owing to the even symmetry of the phonon propagator $D_{\bq \nu}(i\ot_l) = 2\omega_{\bq \nu}/[(i\ot_l)^2 - \omega_{\bq \nu}^2]$ and the resulting structure of the Eliashberg equations, the self-energy components $Z(\ioj)$, $\chi(\ioj)$, and $\phi(\ioj)$ are even functions of the Matsubara index $j$. This allows for efficient summation by exploiting the identity $\omega_{-j} = -\omega_{j-1}$ and regrouping terms with appropriate sign conventions.

As an example, we demonstrate this procedure for the adiabatic contribution to the mass renormalization function $Z^{(\rm A)}(\ioj)$ from Eq.~(37) of the main text. The expression reads:
\begin{align}\label{eq:z_ad-Nc}
Z^{(\rm A)}(\ioj) 
&= 1 + \frac{k_{\rm B}T}{\oj N_F} \sum_{j'=-(N_c+1)}^{N_c} 
\l(\ioj-\iojp)
 \int_{-\infty}^{\infty} d\ve' N(\ve') \g^Z(\ve', \iojp).
\end{align}
Splitting the sum into negative and non-negative components, we get
\begin{align}
Z^{(\rm A)}(\ioj)&= 
1 + \frac{k_{\rm B}T}{\oj N_F} \left [ \sum_{j'=-(N_c+1)}^{-1} 
\l(\ioj-\iojp)
 \int_{-\infty}^{\infty} d\ve' N(\ve') \g^Z(\ve', \iojp)
 \right. \nonumber \\
& \left. \hspace{2cm}+ \sum_{j'=0}^{N_c} 
\l(\ioj-\iojp)
 \int_{-\infty}^{\infty} d\ve' N(\ve') \g^Z(\ve', \iojp) 
 \right ]. 
\end{align}
For the negative indices ($j'< 0$), substituting $j' \to -j'-1$ maps the sum over $j' \in [-N_c - 1, -1]$ to $j' \in [0, N_c]$:
\begin{align}\label{Z_ad_ve_mats-final}
Z^{(\mathrm{A})}(\ioj)
&= 1 + \frac{k_{\rm B}T}{\oj N_F} \sum_{j'=0}^{N_c} \left[
\l(\ioj + \iojp) \int d\ve' N(\ve') \g^Z(\ve', -\iojp)
+ \l(\ioj - \iojp) \int d\ve' N(\ve') \g^Z(\ve', \iojp)
\right] \nonumber \\
&= 1 + \frac{k_{\rm B}T}{\oj N_F} \sum_{j'=0}^{N_c} 
\left[\l(\ioj - \iojp) - \l(\ioj + \iojp) \right]
\int d\ve' N(\ve') \g^Z(\ve', \iojp).
\end{align}

A similar approach can be applied to the Matsubara summations for other quantities, including $\phi^{(\rm A)}(\ioj)$, $\chi^{(\rm A)}(\ioj)$, $N_{\rm e}$ and the terms corresponding to the non-adiabatic vertex corrections. This transformation ensures that all summations are evaluated over positive Matsubara frequencies only, thereby simplifying the numerical implementation.


\section{\label{sec:iso-FSR} Derivation of the isotropic Fermi-surface restricted expressions}

We further simplify the isotropic Eliashberg equations by assuming the constant density of states, i.e., $N(\ve') \rightarrow \NF$. Under this assumption, the adiabatic self-energy in Eq.~\eqref{self-A-iso-simplfied} becomes
\begin{align}\label{self-A-iso-FSR1}
    \hS^{\mathrm{(A)}}(\ioj) 
    &= k_{\rm B} T \sum_{j'} \frac{\l(\ioj-\iojp)}{\NF} \NF 
    \int_{-\infty}^{\infty} 
    d\ve' \hat{\tau}_3 \hG(\ve',\iojp) \hat{\tau}_3
    \nonumber \\
    &= 
    -k_{\rm B} T \sum_{j'}
    \int_{-\infty}^{\infty} d\ve' \,
    \frac{ \iojp Z(\iojp) \hat{\tau}_0 + 
    \left[\ve' -\mu_{\rm F} + \chi(\iojp) \right] \hat{\tau}_3
    - \phi(\iojp) \hat{\tau}_1 } {\left[\o_{j'}Z(\iojp)\right]^2 
    + \left[\ve' -\mu_{\rm F} + \chi(\iojp)\right]^2 
    + \left[\phi(\iojp)\right]^2 }.
\end{align}
We now perform the integral over $\ve'$ analytically. The poles of the Green's function are given by:
\begin{align}\label{ana-int-ve}
    \ve' -\mu_{\rm F} + \chi(\iojp) &=
    \pm i \sqrt{\left[\o_{j'}Z(\iojp)\right]^2 + \left[\phi(\iojp)\right]^2}.
\end{align}
Using the contour integration identity $\int_{-\infty}^{\infty}  \frac{dx}{x^2 + a^2} = \frac{\pi}{a}$ and noting that the integrand term involving $(\ve'-\mu_{\rm F}+\chi)$ is odd in $\ve'$ in the upper-half plane, the relevant integrals reduce to:
\begin{align}\label{eq:analytic-soln-iso}
    & \int d\ve' \frac{1}{{\left[\ojp Z(\iojp) \right]}^2  +
    {\left[ \ve' - \mu_{\rm F} + \chi(\iojp) \right]}^2 + {\left[ \phi(\iojp) \right]}^2} \times \left \{ \begin{array}{c}
         1 \\
         \ve' - \mu_{\rm F} + \chi(\iojp)
    \end{array} \right \} \nonumber \\
    & = \left \{ \begin{array}{c}
         \frac{\pi}{\sqrt{{\left[\ojp Z(\iojp) \right]}^2 
         + {\left[ \phi(\iojp) \right]}^2}} 
         = \frac{\pi}{\Tt(\iojp)}\\
         0 \, ({\rm odd} \, {\rm in} \, \ve'\,)
    \end{array} \right \},
\end{align}
where we define
\begin{subequations}
\begin{align}\label{Th-iso-FSR}
    \Tt(\iojp) &= \sqrt{{\left[\ojp Z(\iojp) \right]}^2 
         + {\left[ \phi(\iojp) \right]}^2}, \\
\tg^Z(\iojp) &= \frac{\ojp Z(\iojp)}{\Tt(\iojp)}, 
\hspace{0.5cm}
\tg^\phi(\iojp) = \frac{\phi(\iojp)}{\Tt(\iojp)}.
\end{align}
\end{subequations}
Substituting these results into Eq.~\eqref{self-A-iso-FSR1}, the adiabatic self-energy under the {\small FSR} limit simplifies to
\begin{align}\label{self-A-iso-FSR}
    \hS^{\mathrm{(A)}}(\ioj) 
    &= -\pi k_{\rm B} T \sum_{j'} \l(\ioj-\iojp)
    \left[i\tg^Z(\iojp) \hat{\tau}_0 - \tg^\phi(\iojp) \hat{\tau}_1 \right].
\end{align}
\par\vspace{1em}
We now turn to the non-adiabatin self-energy in Eq.~\eqref{self-NA-iso-simplified}, again assuming $N(\ve') \rightarrow \NF$, $N(\ve'') \rightarrow \NF$, and $N(\etp) \rightarrow \NF$, with $\mu_{\rm F}$ constant,  the expression becomes
\begin{align} \label{self-NA-iso-FSR}
\hS^{\mathrm{(NA)}}(\ioj)
&=(k_{\rm B}T)^2 \NF \sum_{j'j''} 
\l^V(\ioj-\iojp, \ioj-\iojdp) 
 \nonumber \\
&  \times \int_{-\infty}^{\infty} d\ve'
\int_{-\infty}^{\infty}d(\etp)
\int_{-\infty}^{\infty} d\ve''  
\hat{\tau}_3\hG(\ve',\iojp)\hat{\tau}_3\hG(\etp,\iojtp) \hat{\tau}_3\hG(\ve'',\iojdp)\hat{\tau}_3. 
\end{align}

Performing the analytical integrals over $\ve'$, $\ve''$, and $\etp$ in Eq.~\eqref{self-NA-iso-FSR}, the $\chi$ term vanishes since it is an odd function in energy. The integrand is
solved using a similar approach as described in Eq.~\eqref{eq:analytic-soln-iso}, yielding 
\begin{align} \label{self-NA-iso-FSR2}
\hS^{\mathrm{(NA)}}(\ioj)
&=-\pi^3(k_{\rm B}T)^2 \NF \sum_{j'j''} 
\l^{\rm V}(\ioj -\iojp, \ioj -\iojdp) 
\nonumber \\
& \quad  \times
\left [
-i\tg^Z(\iojp) \tg^Z(\iojtp)\tg^Z(\iojdp) \hat{\tau}_0
+ \tg^Z(\iojp)\tg^Z(\iojtp)\tg^\phi(\iojdp) \hat{\tau}_1 \right. \nonumber \\
&\left. \quad \quad 
-\tg^Z(\iojp)\tg^\phi(\iojtp) \tg^Z(\iojdp) \hat{\tau}_1
-i\tg^Z(\iojp) \tg^\phi(\iojtp)\tg^\phi(\iojdp)\hat{\tau}_0 \right. \nonumber \\
&\left. \quad \quad 
+ \tg^\phi(\iojp) \tg^Z(\iojtp)\tg^Z(\iojdp)\hat{\tau}_1
+ i\tg^\phi(\iojp) \tg^Z(\iojtp) \tg^\phi(\iojdp)\hat{\tau}_0 \right. \nonumber \\
&\left. \quad \quad  
-i\tg^\phi(\iojp) \tg^\phi(\iojtp) \tg^Z(\iojdp) \hat{\tau}_0
+\tg^\phi(\iojp) \tg^\phi(\iojtp)\tg^\phi(\iojdp) \hat{\tau}_1 
\right ] .
\end{align}
Thus, the total self-energy within the isotropic {\small FSR} approximation is the sum of the adiabatic term in Eq.~\eqref{self-A-iso-FSR} and the non-adiabatic term in Eq.~\eqref{self-NA-iso-FSR2}.
Following the standard procedure, we match the coefficients of $\hat{\tau}_0$ and $\hat{\tau}_1$ in the total self-energy. Equating $\hat{\tau}_0$ terms, we obtain the expression for $Z(\ioj)$: 
\begin{align}\label{Z-iso-FSR}
Z(\ioj) = 1 & +
\frac{\pi k_{\rm B} T}{\o_{j}} \sum_{j'} \l(\ioj-\iojp)
 \tg^Z(\iojp) \nonumber \\  
&+ \frac{\pi^3(k_{\rm B}T)^2 \NF}{\o_{j}} \sum_{j'j^{''}}
\l^V(\ioj -\iojp, \ioj -\iojdp)  \nonumber \\
& \quad  \times
\left [
-\tg^Z(\iojp) \tg^Z(\iojtp)\tg^Z(\iojdp) 
-\tg^Z(\iojp) \tg^\phi(\iojtp)\tg^\phi(\iojdp) \right. \nonumber \\
&\left. \quad \quad 
+ \tg^\phi(\iojp) \tg^Z(\iojtp) \tg^\phi(\iojdp)
-\tg^\phi(\iojp) \tg^\phi(\iojtp) \tg^Z(\iojdp) \right ] \nonumber \\
= 1 & + \frac{\pi k_{\rm B} T}{\o_{j}} \sum_{j'} \l(\ioj-\iojp)
 \tg^Z(\iojp) \nonumber \\  
&+ \frac{\pi^3(k_{\rm B}T)^2 \NF}{\o_{j}} \sum_{j'j^{''}} 
\l^V(\ioj -\iojp, \ioj -\iojdp)   \left[ \tg^T(\iojp) \Tilde{P}^Z(\iojtp) \tg(\iojdp) \right].
\end{align}
Similarly, matching $\hat{\tau}_1$ coefficients gives the expression for $\phi(\ioj)$:
\begin{align}\label{phi-iso-FSR}
\phi(\ioj) 
=& 
\, \pi k_{\rm B}T \sum_{j'} 
\l(\ioj-\iojp)\tg^\phi(\iojp) \nonumber \\
&+\pi^3(k_{\rm B}T)^2 \NF \sum_{j'j''} 
\l^{\rm V}(\ioj -\iojp, \ioj -\iojdp)
\nonumber \\
& \quad  \times
\left [
-\tg^Z(\iojp)\tg^Z(\iojtp)\tg^\phi(\iojdp) 
+\tg^Z(\iojp)\tg^\phi(\iojtp) \tg^Z(\iojdp) \right. \nonumber \\
&\left. \quad \quad 
- \tg^\phi(\iojp) \tg^Z(\iojtp)\tg^Z(\iojdp)
-\tg^\phi(\iojp) \tg^\phi(\iojtp)\tg^\phi(\iojdp) \right ]
\\
=& 
\, \pi k_{\rm B} T \sum_{j'} 
\l(\ioj-\iojp)\tg^\phi(\iojp)
\nonumber \\
&+\pi^3(k_{\rm B}T)^2 \NF \sum_{j'j''} 
\l^{\rm V}(\ioj -\iojp, \ioj -\iojdp)
\left [ \tg^T(\iojp) \tP^{\phi}(\iojtp) 
\tg(\iojdp) \right], 
\end{align}
where $\tg$, $\tP^Z$ and $\tP^\phi$ expressions are given by Eqs.~(52)$-$(54) in the main text.



\section{\label{sec:har-conv}Convergence of harmonic and anharmonic phonon calculations for H$_3$S}

\begin{figure}[!t]
    \centering
    \includegraphics[width=0.85\textwidth]{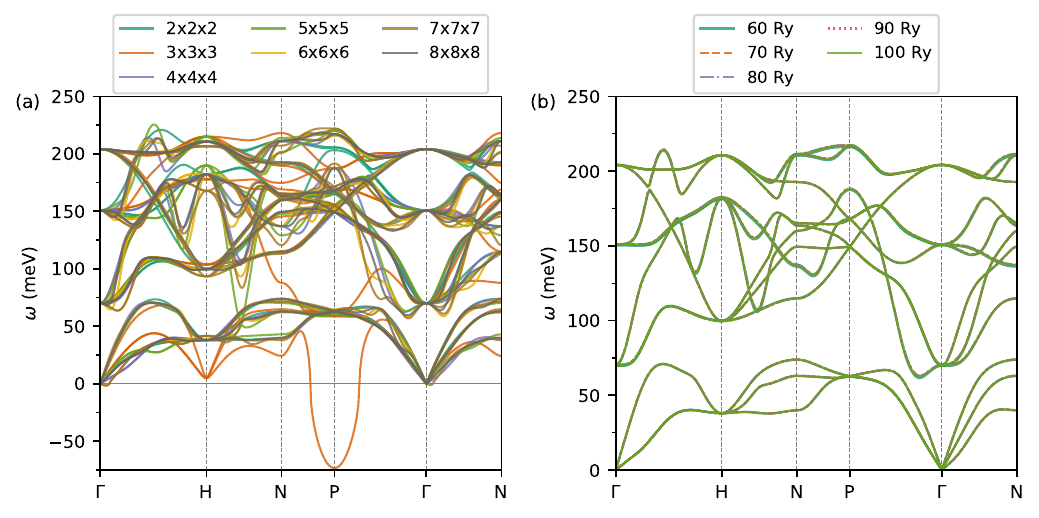}
    \caption{(a) Convergence of the harmonic phonon dispersion of H$_3$S at 200~GPa with respect to the $\bq$-point sampling, ranging from $2 \times 2 \times 2$ to $8 \times 8 \times 8$. Notably, imaginary modes appear when using a $3 \times 3 \times 3$ $\bq$-grid, which are eliminated when anharmonic corrections are included. (b) Convergence of harmonic phonon dispersion with respect to the kinetic energy cutoff, ranging from 60~Ry to 100~Ry. A cutoff of 60~Ry is found to be sufficient for accurate phonon calculations.}
    \label{fig:conv-qgrid-ecut}
\end{figure}

In order to achieve convergence of the harmonic phonons for H$_3$S at 200~GPa, we varied the $\bq$-mesh and computed the phonon dispersions shown in Fig.~\ref{fig:conv-qgrid-ecut}(a). It indicates high sensitivity to the $\bq$-point meshes. For instance, a $3 \times 3 \times 3$ $\bq$-grid results in the emergence of imaginary modes, highlighting the necessity of incorporating anharmonic effects to accurately describe the phonon properties of H$_3$S~\cite{Errea2015,Taureau2024}. Further Fig.~\ref{fig:conv-qgrid-ecut}(b) shows the convergence of phonon dispersion with respect to the kinetic energy cutoff, indicating that a cutoff of 60~Ry is sufficient in this system.

\begin{figure}[!hbt]
    \centering
    \includegraphics[width=0.85\textwidth]{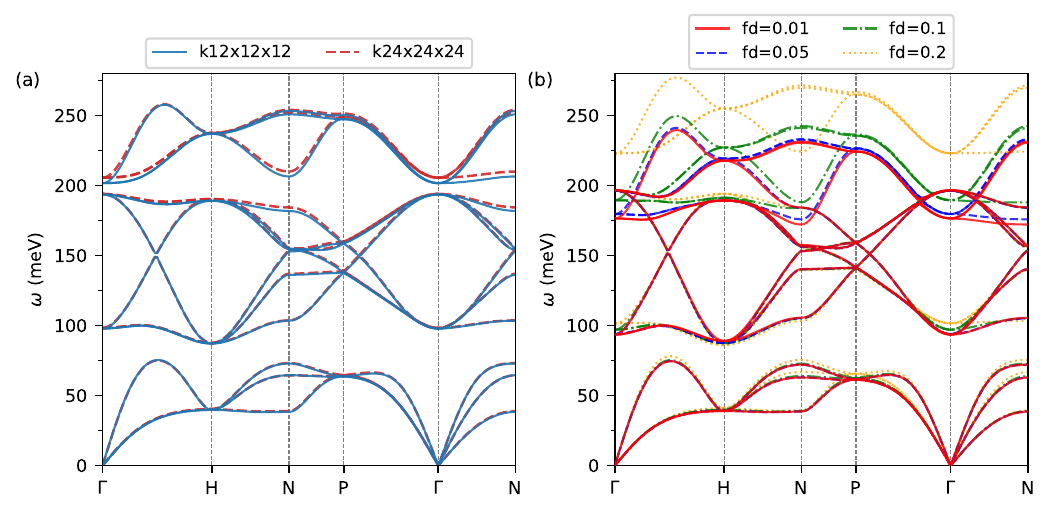}
    \caption{(a) Comparison of anharmonic phonon dispersion of H$_3$S at 200~GPa using the IFCs from the first ZG iteration, computed with a $2 \times 2 \times 2$ supercell and $\bk$-meshes of $12 \times 12 \times 12$ (blue solid line) and $24 \times 24 \times 24$ (red dashed line). 
    (b) Anharmonic phonon dispersion computed using finite displacement amplitudes ranging from 0.01~\AA{} to 0.2~\AA{}. At large displacements (e.g., 0.2~\AA{}), the two highest frequency optical phonon branches are separated from the rest, signaling an unphysical behavior. A displacement of 0.01~\AA{} yields converged and physically meaningful results, and is adopted throughout this study.}
    \label{fig:conv-fd-conv}
\end{figure}
\begin{figure}[!hbt]
    \centering
    \includegraphics[width=0.85\textwidth]{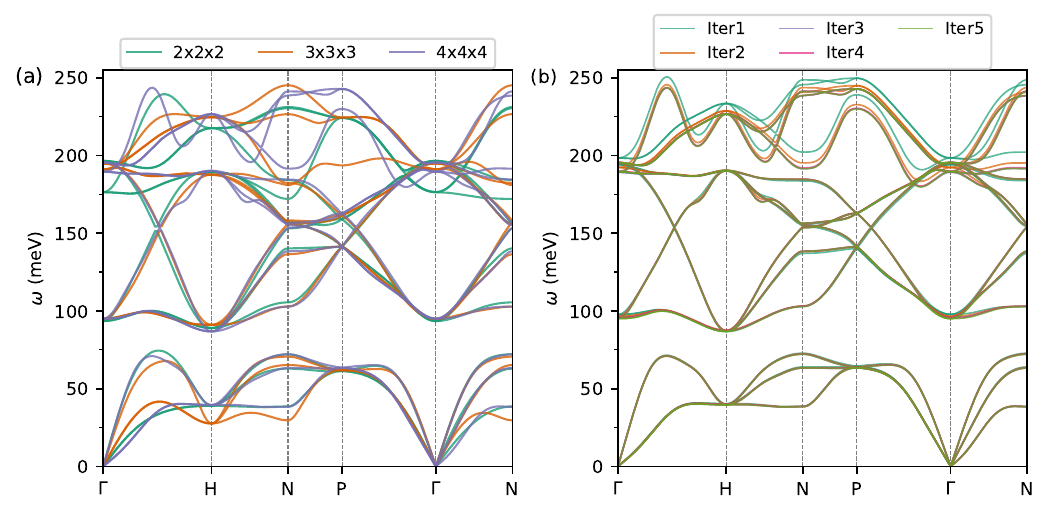}
    \caption{(a) Convergence of anharmonic phonon spectra of H$_3$S at 200~GPa with respect to supercell size, ranging from $2 \times 2 \times 2$ to $4 \times 4 \times 4$.    
    (b) Convergence of the anharmonic phonon spectrum as a function of the number of iterations in the ZG scheme using a linear mixing scheme. IFCs at each iteration were evaluated by finite differences with atomic displacements of 0.01~\AA.} 
    \label{fig:conv-zg}
\end{figure}

We employed the special displacement method implemented in the ZG code~\cite{Zacharias2020,Zacharias2023} to compute phonons with anharmonic corrections. Since these calculations become increasingly expensive with larger supercell sizes, we assessed the phonon convergence using a $2 \times 2 \times 2$ supercell with $\bk$-meshes of $12 \times 12 \times 12$ and $24 \times 24 \times 24$. The close agreement between the results shown in Fig.~\ref{fig:conv-fd-conv}(a) indicates that the coarser $12 \times 12 \times 12$ mesh is sufficient for accurately computing interatomic force constants (IFCs) in the special displacement framework, and is thus adopted in our anharmonic phonon calculations. 
We also examined the convergence of anharmonic phonon spectra with respect to the magnitude of atomic displacements within a $2 \times 2 \times 2$ supercell. As shown in Fig.~\ref{fig:conv-fd-conv}(b), a displacement amplitude of 0.01\AA{} yields converged results and is therefore adopted in this study. Larger displacements lead to unphysical shifts in the optical phonon branches associated with hydrogen vibrations.

Finally, we examined the convergence of anharmonic phonon frequencies using $2 \times 2 \times 2$, $3 \times 3 \times 3$, and $4 \times 4 \times 4$ supercells. As shown in Fig.~\ref{fig:conv-zg}(a), all three supercell sizes yield closely matching results in the low- and mid-frequency ranges. However, the $4 \times 4 \times 4$ supercell better resolves fine features in the high-frequency region, particularly those associated with hydrogen vibrations. For improved accuracy, we therefore adopted the $4 \times 4 \times 4$ supercell in all subsequent anharmonic calculations. Using a linear mixing scheme within the ZG framework, the anharmonic spectrum for H$_3$S typically converged within 4–5 iterations, as illustrated in Fig.~\ref{fig:conv-zg}(b).

\begin{figure}[!t]
    \centering
    \includegraphics[width=0.85\textwidth]{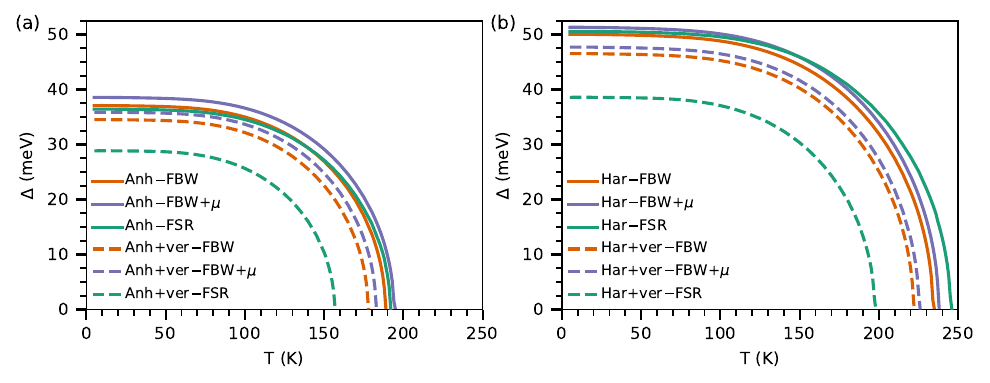}
    \caption{The isotropic superconducting gap $\Delta(T)$ for H$_3$S at 200~GPa computed using various levels of theoretical approaches with an effective Coulomb parameter $\mu_{\rm E}^*$= 0.16 for (a) anharmonic phonons, and (b) harmonic phonons. Solid lines denote adiabatic calculations ({\small FBW}: orange, {\small FBW}+$\mu$: purple, and {\small FSR}: teal-green), while dashed lines include vertex corrections (ver-{\small FBW}: dashed orange, ver-{\small FBW}$+\mu$: dashed purple, and ver-{\small FSR}: dashed teal-green). }
    \label{fig:gap-all}
\end{figure}

\section{\label{sec:gap-fsr}Superconducting properties with vertex corrections for H$_3$S}
Figure~\ref{fig:gap-all}(a) shows the isotropic superconducting gap $\Delta(T)$ for H$_3$S computed using the {\small FSR}, {\small FBW}, and {\small FBW}+$\mu$ methods with anharmonic phonons, both with (dashed lines) and without vertex corrections (solid lines). In the adiabatic {\small FSR} approach, we obtain a critical temperature $\tc\approx$ 192~K, slightly higher than the values obtained with {\small FBW} (189~K) and {\small FBW}+$\mu$ (195~K), due to the omission of the energy dependence of the electronic density of states (DOS). When vertex corrections are included within the {\small FSR} framework, $\tc$ is significantly reduced by approximately 35~K, yielding a value that falls well below the experimental range of 175$-$185~K~\cite{Drozdov2015}. By contrast, including vertex corrections within the {\small FBW} and {\small FBW}+$\mu$ schemes leads to a more moderate 10~K reduction in $\tc$, resulting in predictions that closely match experimental measurements.
\begin{table*}[!t]
    \caption{\label{tab:gap} Zero-temperature superconducting gap $\Delta(0)$, critical temperature $\tc$, and BCS ratio $2\Delta(0)/k_{\rm B} \tc$ computed using the adiabatic ({\small FBW}, {\small FBW}+$\mu$, and {\small FSR}) and vertex-corrected (ver-{\small FBW}, ver-{\small FBW}+$\mu$, and ver-{\small FSR}) approaches with harmonic and anharmonic phonons. A Coulomb pseudopotential of $\mu_{\rm E}^* = 0.16$ is used. Other theoretical studies include FSR results from Errea \textit{et al.}~\cite{Errea2015}, obtained using both harmonic and anharmonic phonons, and average values from the anisotropic FBW and FBW+$\mu$ results of Lucrezi \textit{et al.}~\cite{Lucrezi2024}.}

\setlength\tabcolsep{4pt}
\smallskip
\begin{tabular*}{\textwidth}{@{\extracolsep{\fill}}l l c c c c c c}
\hline\hline \noalign{\vskip 1mm}
\multirow{2}{*}{Phonons} & \multirow{2}{*}{Method} 
& \multicolumn{3}{c}{Present work} 
& \multicolumn{3}{c}{Other theoretical studies} \\
\cline{3-5} \cline{6-8}
& & $\Delta(0)$ (meV) & $\tc$ (K) & $2\Delta(0)/k_{\rm B} \tc$ 
  & $\Delta(0)$ (meV) & $\tc$ (K) & $2\Delta(0)/k_{\rm B} \tc$ \\
\hline
\multirow{3}{*}{Anharmonic}
    & {\small FSR} & 36.43 & 192 & 4.40 & 36.6 & 194 & 4.38 \\
    & {\small FBW} & 37.08 & 189 & 4.55 &       &     &      \\
    & {\small FBW}+$\mu$ & 38.56 & 195 & 4.59 & & & \\
\hline
\multirow{3}{*}{Anharmonic}
    & ver-{\small FSR} & 28.85 & 157 & 4.27 &       &     &      \\
    & ver-{\small FBW} & 34.52 & 178 & 4.50 &       &     &      \\
    & ver-{\small FBW}+$\mu$ & 35.86 & 183 & 4.55 & & & \\
\hline
\multirow{4}{*}{Harmonic}
    & {\small FSR} & 50.57 & 246 & 4.77 & 53.0 & 250 & 4.92 \\
    & {\small FBW} & 50.04 & 235 & 4.94 & 54.4 & 250 & 5.05 \\
    & {\small FBW}+$\mu$ & 51.31 & 238 & 5.00 & 50.0 & 232 & 5.00 \\
\hline
\multirow{3}{*}{Harmonic}
    & ver-{\small FSR} & 38.58 & 198 & 4.52 & & & \\
    & ver-{\small FBW} & 46.52 & 222 & 4.86 & & & \\
    & ver-{\small FBW}+$\mu$ & 47.71 & 226 & 4.90 & & & \\
\hline\hline
\end{tabular*}
\end{table*}
To understand the origin of the difference in $\tc$ between the {\small FSR} and {\small FBW} methods, both with and without vertex corrections, it is useful to examine how DOS enters into each formulation. In the adiabatic limit, the {\small FSR} expressions (first line in Eqs.~(50) and (51) of the main text) do not explicitly depend on $\NF$, whereas the {\small FBW} expressions (first line in Eqs.~(37)$-$(39) of the main text) include $\NF$ in the denominator and integrate over the energy-dependent electronic DOS, thereby capturing its full variation around the Fermi level $\ef$. For H$_3$S, which features a prominent van Hove singularity near $\ef$ with $N_{\rm F} \approx$ 0.32~states/eV/spin/cell, this leads to a slightly lower $\tc$ in {\small FBW} compared to {\small FSR}.
In the non-adiabatic regime, vertex corrections further modify the DOS dependence. In the ver-{\small FSR} formulation, $\NF$ enters explicitly in the numerator (second line in Eqs.~(50) and (51) of the main text), amplifying the impact of a peaked DOS at $\ef$. In contrast, the ver-{\small FBW} formulation includes $\NF^2$ in the denominator and performs a threefold integration over the energy-resolved DOS (second line in Eqs.~(37)$-$(39) of the main text), thereby capturing the detailed DOS variation near $\ef$. 

Figure~\ref{fig:gap-all}(b) presents results analogous to those in Fig.~\ref{fig:gap-all}(a), but using harmonic phonons. Consistent with previous studies~\cite{Errea2015,Sano2016}, we find that harmonic phonons significantly overestimate the $\tc$ across all three methods. When vertex corrections are included in the {\small FSR} approach, $\tc$ is reduced by 48~K, an even larger suppression than that observed in the anharmonic case. In contrast, incorporating the energy dependence of the DOS within the {\small FBW} framework yields a more moderate reduction in $\tc$, on the order of 12$-$13~K, comparable to that found with anharmonic phonons. Table~\ref{tab:gap} summarizes the zero-temperature superconducting gap $\Delta(0)$, critical temperature $\tc$, and BCS ratio $2\Delta(0)/k_{\rm B} \tc$ computed in this study within the adiabatic and non-adiabatic formalisms, using harmonic and anharmonic phonons. These results are compared against previous theoretical predictions based on the Eliashberg formalism~\cite{Errea2015,Lucrezi2024} and recent experimental measurements~\cite{Du2025}.
%

%